\documentclass[pdflatex]{sn-jnl}

\usepackage{graphicx}%
\usepackage{multirow}%
\usepackage{amsmath,amssymb,amsfonts}%
\usepackage{amsthm}%
\usepackage{mathrsfs}%
\usepackage[title]{appendix}%
\usepackage{xcolor}%
\usepackage{textcomp}%
\usepackage{manyfoot}%
\usepackage{booktabs}%
\usepackage{algorithm}%
\usepackage{algorithmicx}%
\usepackage{algpseudocode}%
\usepackage{listings}%
\usepackage{natbib}
\newcommand{\rthis}[1]{\textcolor{black}{#1}}
\usepackage{array}
\usepackage{multirow}

\usepackage{multirow}
\usepackage{graphicx}
\usepackage{subcaption}
\usepackage{amsmath}

\begin{document}

\title[Article Title]{Enhanced Astronomical Source Classification with Integration of Attention Mechanisms and Vision Transformers}


\author[1, 2]{\fnm{Srinadh Reddy} \sur{Bhavanam}}\email{srinadhml99@gmail.com} 

\author[1]{\fnm{Sumohana S} \sur{Channappayya}}\email{sumohana@ee.iith.ac.in}

\author[3]{\fnm{Srijith} \sur{P. K}}\email{srijith@cse.iith.ac.in}

\author[4]{\fnm{Shantanu} \sur{Desai}}\email{shntn05@gmail.com}


\affil[1]{\orgdiv{Department of Electrical Engineering}, \orgname{IIT Hyderabad}, \orgaddress{\street{Kandi}, \city{Hyderabad}, \postcode{502284}, \state{Telangana}, \country{India}}}

\affil[2]{\orgdiv{Department of Physics and Astronomy}, \orgname{Clemson University}, \orgaddress{\street{Kinard Lab of Physics}, \city{Clemson}, \postcode{29634}, \state{South Carolina}, \country{USA}}}

\affil[3]{\orgdiv{Department of Computer Science and Engineering}, \orgname{IIT Hyderabad}, \orgaddress{\street{Kandi}, \city{Hyderabad}, \postcode{502284}, \state{Telangana}, \country{India}}}

\affil[4]{\orgdiv{Department of Physics}, \orgname{IIT Hyderabad}, \orgaddress{\street{Kandi}, \city{Hyderabad}, \postcode{502284}, \state{Telangana}, \country{India}}}


\abstract{Accurate classification of celestial objects is essential for advancing our understanding of the universe. MargNet is a recently developed deep learning-based classifier applied to the Sloan Digital Sky Survey (SDSS) Data Release 16 (DR16) dataset to segregate stars, quasars, and compact galaxies using photometric data. MargNet utilizes a stacked architecture, combining a Convolutional Neural Network (CNN) for image modelling and an Artificial Neural Network (ANN) for modelling photometric parameters. Notably, MargNet focuses exclusively on compact galaxies and outperforms other methods in classifying compact galaxies from stars and quasars, even at fainter magnitudes. In this study, we propose enhancing MargNet's performance by incorporating attention mechanisms and Vision Transformer (ViT)-based models for processing image data. The attention mechanism allows the model to focus on relevant features and capture intricate patterns within images, effectively distinguishing between different classes of celestial objects. Additionally, we leverage ViTs, a transformer-based deep learning architecture renowned for exceptional performance in image classification tasks. We enhance the model's understanding of complex astronomical images by utilizing ViT's ability to capture global dependencies and contextual information. Our approach uses a curated dataset comprising 240,000 compact and 150,000 faint objects. The models learn classification directly from the data, minimizing human intervention. Furthermore, we explore ViT as a hybrid architecture that uses photometric features and images together as input to predict astronomical objects. Our results demonstrate that the proposed attention mechanism augmented CNN in MargNet marginally outperforms the traditional MargNet and the proposed ViT-based MargNet models. Additionally, the ViT-based hybrid model emerges as the most lightweight and easy-to-train model with classification accuracy similar to that of the best-performing attention-enhanced MargNet. This advancement in deep learning will contribute to greater success in identifying objects in upcoming surveys like the Vera C. Rubin Large Synoptic Survey Telescope.}




\maketitle

\section{Introduction}
\label{sec:intro}
Classifying celestial objects is a longstanding challenge in observational astronomy. The segregation between stars, quasars, and galaxies in astronomical images is pivotal to harnessing the best science out of the data. Efficient and robust ways of classification have become of topical importance due to the plethora of ongoing and upcoming astronomical surveys such as the Dark Energy Survey (DES)-~\citep{dark2005dark}, the Sloan Digital Sky Survey (SDSS)-\citep{york2000sloan}, the Zwicky Transient Facility (ZTF)-\citep{bellm2014zwicky}, Euclid~\citep{Euclid}, the Subaru Prime Focus Camera \citep{miyazaki2002subaru}, etc. The upcoming Vera C. Rubin Observatory's LSST \citep{ivezic2019lsst} will further contribute to this data deluge, collecting approximately 15 terabytes of data every night. Given this large volume of data, manual analysis becomes impractical, necessitating the development of automated tools for accurate classification of astronomical sources. These tools are crucial for advancing our understanding of the universe, especially cosmology and structure formation. A more detailed discussion of the impact of source separation on cosmological analyses can be found in ~\citep{Soum}.

Over the past two decades, the evolution of Machine Learning (ML), especially Deep Learning (DL), has significantly impacted astronomy ~\citep{ball2010data, baron2019machine, LahavDL}. ML, with its data-driven training for learning a task, and DL, utilizing multi-layer neural network architectures to acquire hierarchical feature representations, have revolutionized tasks in many areas of astrophysics. A non-exhaustive list of applications of ML to astrophysics includes the classification of stellar spectra~\citep{kuntzer2016stellar, sharma2020application, sharma2020stellar}, photometric light curves ~\citep{lochner2016photometric, mahabal2019machine, moller2020supernnova}, and galaxy morphologies ~\citep{barchi2020machine,gupta2022galaxy}, photometric redshift estimation ~\citep{pasquet2019photometric}, gravitational wave analysis~\citep{george2018deep}, identification of strong lenses ~\citep{cheng2020identifying},  separation of pulsars signals from radio frequency interference~\citep{Bethapudi}, etc. Moreover, DL algorithms have been shown to be very effective in detecting contaminants in astronomical images such as cosmic ray hits~\citep{zhang2020deepcr, xu2023cosmic, bhavanam2022cosmic, bhavanam2022cosmicc}, spurious reflections (``ghosts''), and light scattering~\citep{tanoglidis2022deepghostbusters, chang2021machine}. This integration enhances efficiency and accuracy in various astronomical tasks, showcasing the ongoing potential for data-driven advancements in the field.

The growing availability of data has elevated ML as a powerful tool for astronomical source classification tasks. In the past, methods like Random Forest (RF)-~\citep{vasconcellos2011decision}, Support Vector Machines (SVM)-~\citep{fadely2012star}, and Gradient Boosting classifiers ~\citep{lopez2019j} were favoured for their ease of training with limited data but suffered from weaker performance, typically achieving an accuracy between 80$\%$ and 90$\%$ for the star-galaxy separation. More recent approaches utilizing Convolutional Neural Networks (CNNs)-~\citep{kim2016star, hao2017stacked} have yielded significantly improved results. These advancements were facilitated by the ML community's progress in developing enhanced image classification algorithms, as exemplified by the ImageNet dataset ~\citep{krizhevsky2012imagenet, simonyan2014very}. 

Star-galaxy classification is pivotal to achieving the science requirements from ongoing and next-generation photometric surveys. Although many metrics based on point-spread function (PSF) information have been used for star-galaxy separation~\citep{Desai12,slater2020morphological}, they cannot easily separate stars and quasars. Therefore, many works have explored star-galaxy-quasar separation using ML. While some earlier studies addressed the entire star, quasar and galaxy separation problem using classical ML and DL algorithms ~\citep{Odewahn92, nakazono2021discovery, xiao2021classification, wang2022j}, it is essential to note that there is often a faint limit in source classification, beyond which accurate categorization becomes challenging. For instance, current methods experience performance degradation when dealing with objects \rthis{fainter} than \textit{r} $> 22.5$ ~\citep{cabayol2019pau} and \textit{i} $> 21$ ~\citep{kim2016star}. With the advent of new sky surveys like the Vera Rubin LSST, which observe even fainter celestial objects, achieving accurate classification at these faint magnitudes becomes increasingly paramount.

In this study, we leverage DL to distinguish stars, quasars, and compact galaxies at faint magnitudes, where stars and quasars appear as point sources convolved with the PSF. On the other hand, galaxies are extended sources. Traditionally, differentiating stars and quasars relied on spectra, a resource-intensive process. ML, particularly color-based methods, provides an efficient alternative ~\citep{abraham2018detection} for the star-quasar classification. However, distinguishing high-redshift quasars remains challenging due to intrinsic color variations. Morphological approaches for galaxies~\citep{sebok1986angular} face limitations with fainter and more compact objects, prompting the exploration of spectral energy distributions, although hindered by data availability. Our investigation extends to studying attention-augmented CNNs and Vision Transformer (ViT) architectures, offering potential enhancements for image-based source classification tasks in astronomy and addressing challenges posed by traditional methods.
 
Recently, a novel DL framework called \textit{``MargNet''} \citep{chaini2023photometric} (C23, hereafter) was introduced for classifying stars, quasars and compact galaxies at faint magnitudes with enhanced accuracy. This method employs a DL architecture that combines photometric features-based and image-based classification, using a stacking ensemble of an Artificial Neural Network (ANN) and a CNN, respectively. The model's training dataset incorporates photometric features and images in five photometric bands ($u$, $g$, $r$, $i$, $z$) from the SDSS Data Release (DR)16 ~\citep{ ahumada202016th}. The objects have been classified as stars, quasars or galaxies spectroscopically. Selection criteria outlined in Section 2 of C23 ensured the inclusion of compact galaxies and faint sources by applying constraints on the de Vaucouleur radius and photometric magnitude. 
Their study conducted three experiments to assess the performance in the faint and compact regime, emphasizing the importance of precise source separation.

This work addresses the classification challenge of faint and compact astronomical objects by enhancing the ``MargNet'' model architecture initially used in C23. We extend our investigation to incorporate attention-enhanced CNNs and ViT architectures to better handle image data. In particular, we integrate the Squeeze and Excitation (SE)-\citep{hu2018squeeze} attention module into CNNs, a technique that has previously shown substantial performance improvements in image classification tasks. \rthis{Therefore, this work is an extension of the analysis carried out in C23, where we use the same dataset and analysis cuts, ensuring that the evaluation of the new enhancements remains consistent and comparable to the previous findings.}


First introduced by ~\citep{dosovitskiy2020image}, ViTs excel in image classification, object detection~\citep{li2022exploring}, and semantic segmentation~\citep{chen2021transunet} tasks. ViTs represent a revolutionary paradigm in computer vision, diverging from traditional CNNs by adopting self-attention mechanisms. Their unique approach to processing image patches through self-attention allows them to capture global dependencies and intricate spatial relationships, leading to state-of-the-art performance. ViTs have also proven effective in transfer learning, particularly in scenarios with limited labelled data, and found applications in medical image analysis~\citep{he2023transformers}, remote sensing~\citep{bazi2021vision}, and various domains where understanding global context is critical. 

While attention-enhanced CNNs have improved performance over baseline CNN models, their application in astronomy remains relatively unexplored, except for contaminant detection~\citep{bhavanam2022cosmic}. Although ViTs have been investigated for galaxy morphology classification ~\citep{karpoor2022morphological, lin2021galaxy}, their potential to address the source separation problem remains untapped. Further, by leveraging the capabilities of ViT, we aim to extract and analyze intricate features from diverse astronomical data types, including photometric features and FITS images. This approach facilitates the identification of unique signatures associated with various classes of celestial objects.

Our primary contributions in this work are described as follows:

\begin{itemize}
    \item Exploring the modality-specific source classification models tailored to handle photometric features and FITS images separately.
    \item Investigating the optimal approach to source classification using FITS images solely, comparing the performance of CNNs with and without attention augmentation and ViT-based models.
    \item Unraveling the ``MargNet'' model and expanding its capabilities by integrating attention-enhanced CNNs and ViT models for the image-based classifier. 
    \item Exploring the synergistic potential of integrating photometric features and FITS images through fusion in ViT-based architectures. This augmentation aims to enhance astrophysical source detection capabilities within the baseline ViT framework.
\end{itemize}

The remainder of this paper is structured as follows: Section \ref{sec:dataset} covers data acquisition from the SDSS DR16 Catalogues. Section \ref{sec:methods} describes the preprocessing steps, the methods employed for the classification problem and details on the models. In Section \ref{sec:results}, we present the results and discussion; in Section \ref{sec:conclusion}, we present the conclusions and further possibilities arising from this work.

\section{Dataset and Experiments}
\label{sec:dataset}
The primary objective of this study is to reliably classify the astrophysical sources with faint and compact characteristics. 
For this study, we rely on SDSS DR 16~\citep{ahumada202016th}, specifically utilizing SDSS photometric data in the $u$, $g$, $r$, $i$, and $z$ passbands~\citep{fukugita1996sloan} as model inputs. Spectroscopically assigned classes from the SDSS pipeline serve as labels for supervised training, ensuring the accuracy of our model predictions. We use both photometric features and FITS images from the SDSS DR 16 for our analysis.

\subsection{Compact  and Faint source dataset}
\label{sec:faint_copmact_sec}
To create a sample of galaxies which can be identified as compact, we use the same compactness parameter ($\textit{c}$) introduced in C23 and choose the threshold of $c=0.5$ to separate the two.
To constitute the faint object dataset, we impose the following constraint on  the average magnitude across the five passbands from SDSS data, similar to C23:
\begin{equation} 
\label{eq:eq2}
\langle mag \rangle > 20
\end{equation}

\rthis{More detailed discussion of these cut choices can be found in C23.}
Using these cuts, we create both a Compact source dataset as well as a Faint and Compact source dataset using DR16 data in the same way as C23.

After retrieving the photometric features and FITS images from both the datasets, we are set up to conduct the experiments. We split the dataset into training, validation, and test sets, ensuring an equal representation of each class. We prepared three different experiments, which we label as \textbf{Experiment 1}, \textbf{Experiment 2}, and \textbf{Experiment 3} and the details of each of these experiments can be found below:

\begin{itemize}
    \item \textbf{Experiment 1:} We obtain all three sets: training, validation, and test from the Compact source dataset. Importantly, the test set is representative of the training set in this context.
    \item \textbf{Experiment 2:} The training, validation, and test sets are drawn from the Faint and Compact source dataset.
    \item \textbf{Experiment 3:} The training and validation sets are selected from the Compact source dataset. However, the test set is chosen from the Faint and Compact source dataset.
\end{itemize}
\rthis{Further details about these experiments on data splitting are exactly the same as C23,  and we omit the details for brevity. } 

\section{Methodology}
\label{sec:methods}
The primary objective of this work is to explore optimal models for source classification, utilizing both the photometric features and FITS images. Our study includes the independent and joint analysis of models leveraging these two types of features. 

\subsection{Classification using Photometric Features}
We utilized 24 distinct features, as outlined in Table 1 of C23, to represent the photometric information derived from celestial source images. This phase capitalizes on the wealth of photometric features to extract insights into celestial objects, encompassing brightness, color, and variability. The classification task using photometric features involves employing classical supervised ML classifiers alongside ANNs. ML classifiers, including Decision Trees (DT), RF, and XGBoost (XGB), play a pivotal role in classifying entities such as star-galaxy-quasars. Furthermore, adhering to the model proposed in C23, incorporating ANN classifiers offers an additional approach for achieving precise and accurate classification of celestial objects.

\textbf{1. Decision Trees}, a prevalent ML technique in astronomy for tasks like classification and regression, operate by recursively partitioning the data based on feature conditions and establishing a tree ~\citep{quinlan1986induction}. Each split in the tree represents a decision, and collectively, they form a hierarchical structure. DTs are known for their interpretability and simplicity, enabling clear insights into decision-making. However, they may be prone to overfitting and capturing noise in the training data. Regularization techniques, ensemble methods like RF, and careful tuning help mitigate these issues, making DTs valuable in astronomical analysis ~\citep{vasconcellos2011decision}.

\textbf{2. Random Forest}, a widely used ML technique in astronomy ~\citep{baron2019machine}, creates ensembles of DTs through bootstrap resampling, preventing overfitting by leveraging diverse subsets of features during training. Introduced in ~\citet{breiman2001random}, RF excels in efficiency, scalability, and competitive performance for classification and regression in astronomy compared to other ML algorithms.

\textbf{3. XGBoost (XGB)}, an influential ML algorithm widely applied in astronomy for tasks such as classification and regression ~\citep{Bethapudi,zhang2022classification}, stands out for its exceptional predictive performance. An optimized gradient boosting framework, XGB sequentially builds a strong predictive model by combining weak learners (typically DTs). It incorporates regularization techniques to control model complexity and prevent overfitting ~\citep{chen2016xgboost}, and it includes advanced features like parallel processing and handling missing values. Renowned for its speed, efficiency, and ability to handle large datasets, XGBoost has become popular in astronomy for achieving accurate and robust predictions.

\textbf{4. Artificial Neural Networks}, inspired by biological neural networks, ANNs emulate biological neurons, receiving inputs, processing information, and producing outputs. The ANN model in this study, (same as in  C23), is a feed-forward network with stacked dense layers activated by ReLU units. Five layers provide ample depth for directly classifying astronomical sources from selected parameters. Dropouts ~\citep{srivastava2014dropout} with a 0.25 fraction are employed to combat overfitting in deep networks. The final layer, softmax-activated with three (or two) outputs for each category, utilizes categorical (or binary) cross-entropy as the loss function for the star-quasar-galaxy classification. 

\subsection{Classification Using FITS Images}
\label{sec:cls_images}
The SDSS FITS images in all five passbands are preprocessed in the same way as C23. We then apply the following methods to these images for astronomical source separation.

\begin{enumerate}
\item  \textbf{Convolutional Neural Networks} are essential for image processing in computer vision. Their strength lies in specialized layers like convolutional and pooling layers, effectively capturing hierarchical image features. CNNs excel at recognizing patterns, edges, and textures, enabling detailed visual analysis. Parameter sharing in convolutional layers reduces model complexity, enhancing generalization across image regions. This adaptability makes CNNs robust for object recognition, image classification, and segmentation. Additionally, CNNs have been successfully applied to astronomical source classification ~\citep{kim2016star}.

\begin{figure}
\centering
\includegraphics[width=0.9\linewidth]{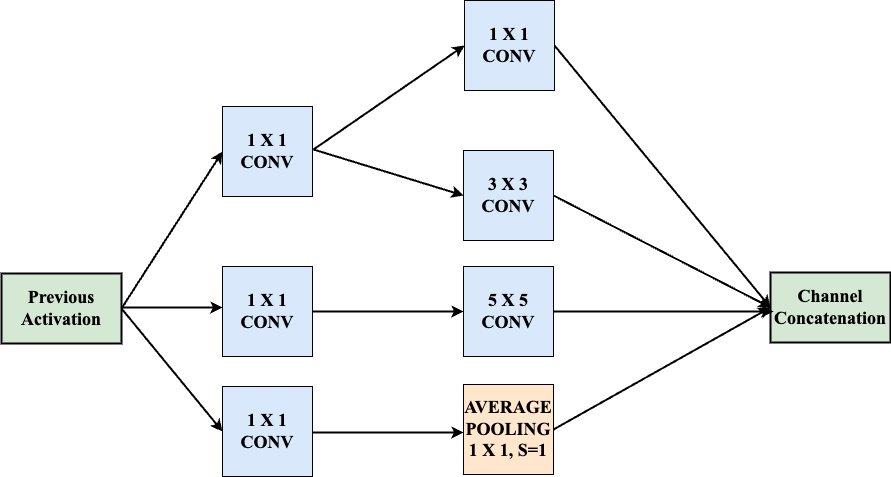}
\caption{The Inception module architecture used.}
\label{fig:inc}
\end{figure}

In C23, a CNN architecture inspired by the Inception Network, as introduced by \citet{szegedy2015going} for the ImageNet Challenge, is employed for compact and faint astronomical source separation. This adaptation of InceptionNet, previously successful in astronomy, utilizes layer sizes from \citet{pasquet2019photometric}, initially designed for photometric redshift estimation.

Including the inception module is crucial, aiding in extracting discriminative features from various convolution filter sizes (1$\times$1, 3$\times$3, and 5$\times$5). These filters are applied at a specific depth, and their outputs are concatenated to create a consolidated layer. We considered the same CNN model from C23 as our baseline and tried to improve the performance over the baseline models across all three experiments. Figure \ref{fig:inc} illustrates the chosen inception module in this work. CNN processes pre-processed FITS image files for data objects, providing initial classification output. With 80,000 images per class and spectroscopic labels, the CNN uses convolutional layers with varying kernel sizes ([1,1], [3,3], [5,5]) activated by ReLU units. Dense layers distil information, and the final layer employs softmax (or sigmoid) for multi-class (or binary) classification. It uses categorical cross-entropy as the loss function.  The complete details of the CNN architecture used can be found in Table 3 of  C23.

\item \textbf{Attention CNN:} As discussed in Section~ \ref{sec:intro}, attention mechanisms play a pivotal role in computer vision in enhancing the capability of models to focus on relevant information within an image~\citep{guo2022attention}. Analogous to human visual attention, these mechanisms enable models to selectively emphasize specific regions or features during processing. By dynamically assigning weights to different input parts, attention mechanisms improve the model's ability to capture intricate patterns, relationships, and contextual information. Integrating attention mechanisms has become a fundamental aspect of modern computer vision architectures, contributing to overall advancement and adaptability in image-based tasks. Furthermore, the attention blocks can be effortlessly incorporated into pre-existing CNN models, allowing for seamless integration without requiring extensive modifications to the architecture.

Inspired by the success of attention-enhanced CNN models, we experimented with various attention mechanisms to enhance the performance of our baseline CNN model for celestial object classification. In particular, we investigated the SENet ~\citep{hu2018squeeze}. We also explored the Simple Attention Module (SimAM; ~\citep{yang2021simam}) and Convolutional Block Attention Module \rthis{(CBAM; ~\citep{woo2018cbam})}. The SENet introduces a mechanism to adaptively recalibrate the importance of different channels within feature maps. The ``squeeze'' operation captures global information by applying global average pooling, while the ``excitation'' operation learns channel-wise dependencies. This enables SENet to dynamically assign weights to channels based on relevance, enhancing the network's discriminative power and performance in tasks such as image classification and object detection. The block diagram representing operations in SENet is illustrated in Figure \ref{fig:se}.

SE networks have emerged as a powerful enhancement to CNNs for image classification tasks. They aim to improve model performance by explicitly modelling channel-wise dependencies, allowing the network to adaptively recalibrate the importance of different channels. One crucial parameter in SE networks is the reduction ratio, often denoted as \( r \). This parameter controls the dimensionality of the intermediate representations within the SE block. Specifically, during the squeeze phase, which involves global average pooling to obtain channel-wise statistics, the reduction ratio determines the number of intermediate features used to capture channel dependencies. Mathematically, if the input tensor has \( C \) channels, the reduction ratio (\( r \)) reduces the number of channels to \( \frac{C}{r} \) before applying the excitation phase. Typically, \( r \) is chosen as a hyperparameter, and commonly used values range from 2 to 16, although smaller values like 2 or 4 are prevalent.

Choosing an appropriate reduction ratio is crucial for balancing model complexity and performance. A higher reduction ratio decreases computational cost but may limit the model's capacity to capture complex channel interactions. Conversely, a lower reduction ratio may improve expressiveness but at the expense of increased computational overhead. The reduction parameter is often determined through experimentation and validation on the specific dataset and task. It is essential to balance model capacity and efficiency to achieve optimal performance.

\begin{figure}
\centering
\includegraphics[width=0.9\linewidth]{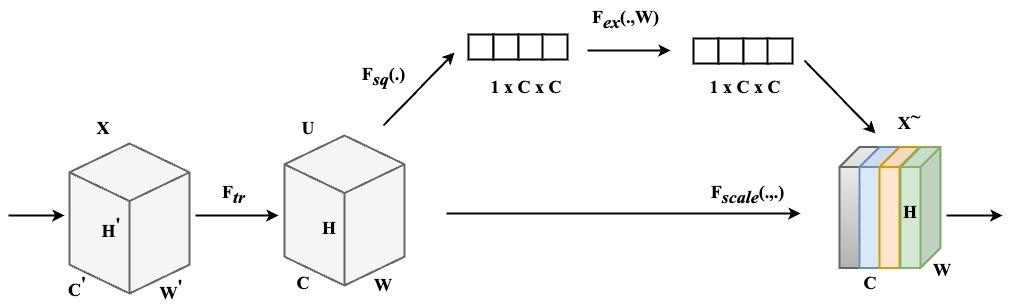}
\caption{The block diagram of the SENet used ~\citep{hu2018squeeze}.}
\label{fig:se}
\end{figure}

\item \textbf{Vision Transformers} represent a paradigm shift in computer vision, leveraging transformer architectures from ~\citet{vaswani2017attention} for image classification tasks ~\citep{dosovitskiy2020image}. Unlike traditional CNNs, ViTs treat images as sequences of patches, allowing them to capture local and global contextual information. This approach enables ViTs to scale efficiently to large image sizes, contributing to their success in various image classification challenges. ViTs have demonstrated state-of-the-art performance on benchmarks such as ImageNet, showcasing their potential to outperform traditional CNNs in certain scenarios. On a related note, Linformer ~\citep{wang2020linformer} is a variant of the ViT architecture that specifically addresses the quadratic complexity of the self-attention mechanism in ViT, making it more computationally efficient. While ViTs have excelled in image classification tasks, Linformer's advancements in attention mechanisms within ViTs make it particularly relevant for applications requiring the processing of extensive image datasets. ViTs and Linformer demonstrate the adaptability of transformer-based architectures in handling diverse image-related tasks, significantly contributing to the evolution of machine learning methodologies in computer vision.

The Classification (CLS) token in the ViT architecture is a pivotal element at the model's beginning. It serves as a representative summary of the entire input image, capturing global contextual information. Traditionally initialized randomly during training, some approaches consider deriving the CLS token intentionally from specific features. This is studied specifically when multimodal data is available in remote sensing applications ~\citep{roy2023multimodal}. This token is crucial in aggregating essential information for downstream tasks, contributing significantly to ViT's visual recognition and understanding performance. In Figure~\ref{fig:mmvit}, the architecture of ViT is illustrated, and the CLS token is extracted randomly while working with images solely.
\end{enumerate}

\begin{figure}
\centering
\includegraphics[width=0.8\linewidth]{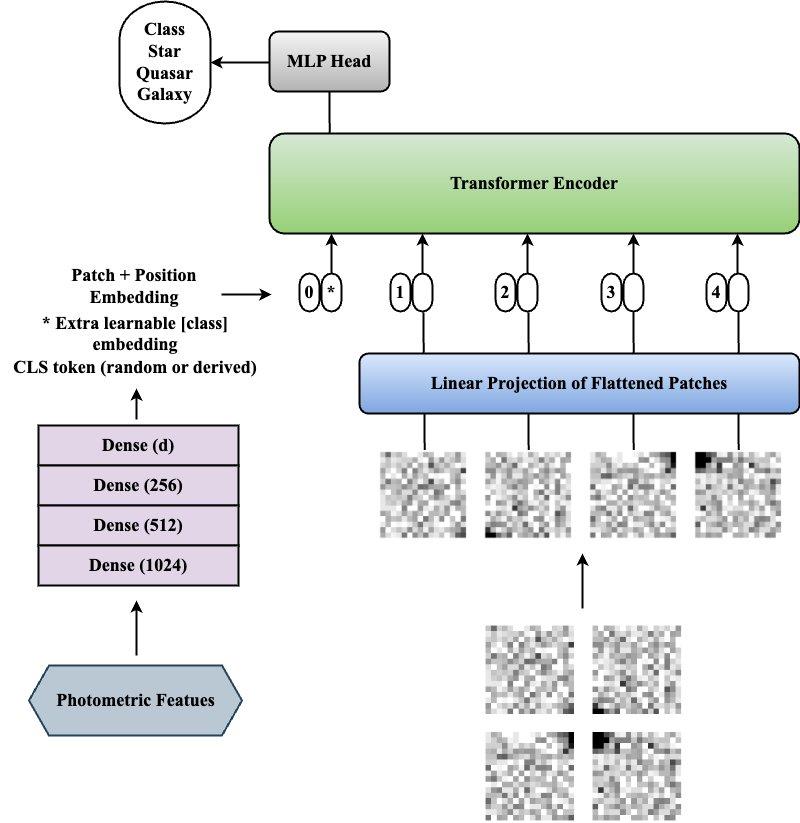}
\caption{The ViT architecture, with its randomly chosen CLS token, represents an image-based model. Conversely, deriving the CLS token using photometric features represents a novel architectural approach, seamlessly fusing photometric features and images.}
\label{fig:mmvit}
\end{figure}

\subsection{Classification Using Photometric Features and FITS Images}
We systematically explored hybrid models that combine photometric features with images, investigating various integration approaches. Our experimentation involved deploying stacking ensemble models, similar to MargNet, which were trained separately using photometric features and images. Later, these models were combined using ensemble techniques. The second idea is to implement nuanced adjustments within the ViT architecture (proposed). This included training a unified model that leverages both photometric features and images. Specifically, we adopted the multi-modal fusion-based Vision Transformer (MM ViT) paradigm, inspired by the insights introduced by ~\citet{roy2023multimodal}. This comprehensive investigation aimed to optimize the fusion of visual and photometric information to enhance model performance. Further details on these approaches are provided below.

\begin{enumerate}
\item 
\textbf{Ensemble Model (MargNet):} The CNN (for images) and ANN (for photometric features) models were trained using an early stopping criterion. These pre-trained models were then integrated in parallel through stacking to create the final network, MargNet. The models were fine-tuned for their inputs, eliminating the need for retraining MargNet models. CNN and ANN outputs, generated from images and photometric data of each class, were used as inputs for the ensemble, \rthis{in the same way as C23, where more details can be found}.  We also explored alternative architectures for MargNet by replacing the CNN model with attention-augmented CNNs and ViTs, maintaining the ANN model unchanged. 

\item 
\textbf{ MultiModal ViT (MM ViT):} This represents a pioneering approach in computer vision by seamlessly integrating visual information from diverse modalities ~\citep{roy2023multimodal}. Drawing inspiration from ViT's success in image classification, MM ViT extends its capabilities to concurrently handle various image modalities, such as multi-spectral and hyper-spectral data. Unlike traditional models that process each modality separately, MM ViT employs a unified architecture, enabling joint learning and information fusion. This framework empowers the model to capture intricate relationships and dependencies across different modalities, leading to more robust and holistic representations. By leveraging self-attention mechanisms and hierarchical feature extraction, MM ViT showcases the potential for enhanced performance in tasks requiring synthesising information from various sources, making it a versatile and powerful tool in multimodal learning scenarios.

Our study treated photometric features and images as distinct modalities and conducted experiments to fuse them seamlessly. We strategically designated the CLS token from the ViT architecture as a pivotal hyperparameter to achieve this. Traditionally, the CLS token in ViT training is selected randomly. However, our methodology advocates deriving the CLS token from photometric features. Simultaneously, the model receives images as input in patch form, facilitating the comprehensive incorporation of both data types. This approach contributes to a more refined and comprehensive understanding of the combined visual and photometric information. The architecture of the proposed MM ViT is illustrated in Figure \ref{fig:mmvit}. The CLS token is extracted by passing the photometric features through a simple ANN model. The final dense layer (with \textit{d} neurons) dimension is adjusted to match the dimension of the flattened image patches. Finally, the CLS token is concatenated with the projected and flattened patches before being fed into the ViT encoder.
\end{enumerate}

\subsection{Performance Evaluation}
To assess the effectiveness of classification models, it is essential to utilize three cornerstone metrics: Precision, Recall, and Accuracy. Precision measures the proportion of correctly identified positive instances (TP) out of all instances classified as positive (TP + FP), providing insight into the model's ability to avoid false positives (FP). Conversely, to prove insight into the model's ability to avoid false negatives (FN), Recall evaluates the model's capability to capture all positive instances by calculating the ratio of true positives (TP) to the total number of positive instances (TP + FN). Accuracy, another crucial metric, measures the accuracy of the model's predictions by calculating the ratio of correctly classified instances (TP + TN) to the total number of instances. Together, these metrics offer a comprehensive understanding of a model's performance across different aspects of classification tasks, thereby facilitating informed decision-making in model selection and optimization.

For a specific class $c$, precision ($P_c$) and recall ($R_c$) are calculated as follows:
\begin{equation}
P_c = \frac{TP_c}{TP_c + FP_c}, \quad R_c = \frac{TP_c}{TP_c + FN_c}, \quad 
\end{equation}
\text{where } c \text{ denotes objects for that particular class.}

We compute the aggregate precision ($P$) and recall ($R$) by averaging the precision ($P_c$) and recall ($R_c$) values across all classes. Precision for galaxies signifies the classifier's proficiency in avoiding misclassifying stars/quasars as galaxies. In contrast, recall for galaxies denotes the classifier's effectiveness in identifying all galaxies in the dataset.

Accuracy can be calculated for a single class or all classes combined together, and is given by:
\begin{equation}
\text{Accuracy} = \frac{TP + TN}{TP + TN + FP + FN}, \quad 
\end{equation}
where $TN$ denotes the true negatives.

\section{Results and Discussion}
\label{sec:results}
For each experiment described in Section \ref{sec:faint_copmact_sec}, we investigate two scenarios: star-galaxy classification and star-galaxy-quasar classification. Subsequent sections delve into the specifics of each methodology, focusing on the corresponding data type and the quantitative performance metrics. We emphasize the results obtained from models utilizing photometric features and FITS images as input. 

\subsection{Performance Analysis with Photometric Features}
The photometric features extracted from the source images exclusively serve as inputs for models that do not incorporate image data. We utilized classical ML algorithms, including DT, RF, and XGB, alongside an ANN model to accommodate these explicit features. 
\rthis{We implemented 10-fold cross-validation while implementing these algorithms to ensure robust classification of both star-galaxy and star-galaxy-quasar classification. Errors and performance metrics were computed for each fold, with averages and errors obtained from standard deviations reported.} Since this analysis does not consider images, no additional preprocessing is necessary apart from standard data normalization. The performance of each model is summarized in Table \ref{tab:results_pf}, and our deep learning models are implemented using the {\tt PyTorch} framework. Table \ref{tab:results_pf} shows that XGB consistently outperforms DT and RF classifiers across all experiments and classification settings. The best performance is achieved by either XGB or ANN. For the ANN model, we employed a dropout fraction of 0.25, and the model comprises 331,331 neurons for the architecture employed.
\begin{table}[h]
\caption{Quantitative performance of the classical ML algorithms and ANN model with photometric features. Models in italics represent re-trained models from C23 in the PyTorch framework. The best-performing models are marked in bold. \rthis{The optimized parameters for the DT classifier are; \texttt{criterion}: \texttt{gini}, \texttt{max\_depth}: 10, \texttt{min\_samples\_split}: 10. The RF has \texttt{max\_depth}=50, \texttt{min\_samples\_split}=5. For GBDT, \texttt{max\_depth}=50, \texttt{n\_estimators}=100. While employing DT, RF, GBDT and ANN models, we utilized 10-fold cross-validation. Performance metrics were averaged across test folds, with errors estimated from the standard deviation.}}
\label{tab:results_pf}
\begin{tabular}{@{}lcccccc@{}}
\toprule
Experiment & Classes & Model & Accuracy ($\%$) & Precision ($\%$) & Recall ($\%$) \\
\midrule
Experiment 1 & 1. Star-Galaxy & DT & \rthis{96.9 $\pm$ 0.1} & \rthis{96.9 $\pm$ 0.1} & \rthis{96.9 $\pm$ 0.1} \\
& & RF & \rthis{97.7 $\pm$ 0.1} & \rthis{97.7 $\pm$ 0.1} & \rthis{97.7 $\pm$ 0.1} \\
& & XGB & \textbf{\rthis{98.0 $\pm$ 0.1}} & \textbf{\rthis{98.0 $\pm$ 0.1}} & \textbf{\rthis{98.0 $\pm$ 0.1}} \\
& & \textit{ANN} & \rthis{\textit{97.9 $\pm$ 0.1}} & \rthis{\textit{97.9 $\pm$ 0.1}} & \rthis{\textit{97.9 $\pm$ 0.1}} \\
& 2. Star-Galaxy-Quasar & DT & \rthis{89.5 $\pm$ 0.2} & \rthis{89.6 $\pm$ 0.2} & \rthis{89.5 $\pm$ 0.2} \\
& & RF & \rthis{92.0 $\pm$ 0.1} & \rthis{92.1 $\pm$ 0.1} & \rthis{92.0 $\pm$ 0.1} \\
& & XGB & \rthis{92.7 $\pm$ 0.2} & \rthis{92.7 $\pm$ 0.2} & \rthis{92.7 $\pm$ 0.2} \\
& & \textit{ANN} & \rthis{\textbf{\textit{92.9 $\pm$ 0.1}}} & \rthis{\textbf{\textit{92.9 $\pm$ 0.1}}} & \rthis{\textbf{\textit{92.9 $\pm$ 0.1}}} \\
\midrule
Experiment 2 & 1. Star-Galaxy & DT & \rthis{94.7 $\pm$ 0.2} & \rthis{94.7 $\pm$ 0.2} & \rthis{94.7 $\pm$ 0.2} \\
& & RF & \rthis{96.1 $\pm$ 0.2} & \rthis{96.1 $\pm$ 0.2} & \rthis{96.1 $\pm$ 0.2} \\
& & XGB & \textbf{\rthis{96.3 $\pm$ 0.1}} & \textbf{\rthis{96.3 $\pm$ 0.1}} & \textbf{\rthis{96.3 $\pm$ 0.1}} \\
& & \textit{ANN} & \rthis{\textit{96.2 $\pm$ 0.2}} & \rthis{\textit{96.2 $\pm$ 0.2}} & \rthis{\textit{96.2 $\pm$ 0.2}} \\
& 2. Star-Galaxy-Quasar & DT & \rthis{81.9 $\pm$ 0.3} & \rthis{82.0 $\pm$ 0.3} & \rthis{81.9 $\pm$ 0.3} \\
& & RF & \rthis{85.5 $\pm$ 0.2} & \rthis{85.6 $\pm$ 0.2} & \rthis{85.5 $\pm$ 0.3} \\
& & XGB & \textbf{\rthis{86.1 $\pm$ 0.2}} & \textbf{\rthis{86.1 $\pm$ 0.2}} & \textbf{\rthis{86.1 $\pm$ 0.2}} \\
& & \textit{ANN} & \rthis{\textit{86.3 $\pm$ 0.2}} & \rthis{\textit{86.4 $\pm$ 0.2}} & \rthis{\textit{86.3 $\pm$ 0.2}} \\
\midrule
Experiment 3 & 1. Star-Galaxy & DT & \rthis{88.9 $\pm$ 0.3} & \rthis{90.3 $\pm$ 0.2} & \rthis{88.9 $\pm$ 0.3} \\
& & RF & \rthis{90.5 $\pm$ 0.2} & \rthis{91.9 $\pm$ 0.1} & \rthis{90.4 $\pm$ 0.1} \\
& & XGB & \rthis{92.0 $\pm$ 0.1} & \rthis{93.1 $\pm$ 0.1} & \rthis{92.0 $\pm$ 0.1} \\
& & \textit{ANN} & \rthis{\textbf{\textit{92.1 $\pm$ 0.8}}} & \rthis{\textbf{\textit{92.9 $\pm$ 0.6}}} & \rthis{\textbf{\textit{92.1 $\pm$ 0.8}}} \\
& 2. Star-Galaxy-Quasar & DT & \rthis{68.9 $\pm$ 0.3} & \rthis{72.8 $\pm$ 0.2} & \rthis{68.9 $\pm$ 0.3} \\
& & RF & \rthis{83.1 $\pm$ 0.6} & \rthis{83.4 $\pm$ 0.6} & \rthis{83.1 $\pm$ 0.6} \\
& & XGB & \textbf{\rthis{83.8 $\pm$ 0.5}} & \textbf{\rthis{83.9 $\pm$ 0.5}} & \textbf{\rthis{83.8 $\pm$ 0.5}} \\
& & \textit{ANN} & \rthis{\textit{72.2 $\pm$ 1.8}} & \rthis{\textit{74.3 $\pm$ 2.0}} & \rthis{\textit{72.1 $\pm$ 1.8}} \\
\bottomrule
\end{tabular}
\end{table}

\begin{table}[h]
\caption{Quantitative performance of the proposed models using FITS images. Models in italics represent the re-trained CNN model from C23, and bold are the best performed. While using SENet added CNN, the hyperparameter `r' for each experiment and class setting is mentioned in the brackets.}
\label{tab:results_images}
\begin{tabular}{@{}lccccc@{}}
\toprule
Experiment & Classes & Model & Accuracy ($\%$) & Precision ($\%$) & Recall ($\%$) \\
\midrule
Experiment 1 & 1. Star-Galaxy & \textit{CNN} & \textit{97.4 $\pm$ 0.1} & \textit{97.4 $\pm$ 0.1} & \textit{97.4 $\pm$ 0.1} \\
& & SENet-CNN & \textbf{97.5 $\pm$ 0.1} & \textbf{97.5 $\pm$ 0.1} & \textbf{97.5 $\pm$ 0.1} \\
& & (r=2) &  &  &  \\
& & ViT & 97.1 $\pm$ 0.1 & 97.1 $\pm$ 0.1 & 97.1 $\pm$ 0.1 \\
& 2. Star-Galaxy-Quasar & \textit{CNN} & \textit{91.6 $\pm$ 0.1} & \textit{91.7 $\pm$ 0.1} & \textit{91.6 $\pm$ 0.1} \\
& & SENet-CNN  & \textbf{91.9 $\pm$ 0.1} & \textbf{92.0 $\pm$ 0.1} & \textbf{91.9 $\pm$ 0.1} \\
& & (r=32) &  &  &  \\
& & ViT & 91.2 $\pm$ 0.1 & 91.4 $\pm$ 0.1 & 91.2 $\pm$ 0.1 \\
\midrule
Experiment 2 & 1. Star-Galaxy & \textit{CNN} & \textit{95.2 $\pm$ 0.1} & \textit{95.2 $\pm$ 0.1} & \textit{95.2 $\pm$ 0.1} \\
& & SENet-CNN & \textbf{95.6 $\pm$ 0.1} & \textbf{95.6 $\pm$ 0.1} & \textbf{95.6 $\pm$ 0.1} \\
& & (r=16) &  &  &  \\
& & ViT & 94.2 $\pm$ 0.1 & 94.3 $\pm$ 0.1 & 94.2 $\pm$ 0.1 \\
& 2. Star-Galaxy-Quasar & \textit{CNN} & \textit{84.2 $\pm$ 0.1} & \textit{84.5 $\pm$ 0.1} & \textit{84.2 $\pm$ 0.1} \\
& & SENet-CNN & \textbf{84.8 $\pm$ 0.1} & \textbf{85.0 $\pm$ 0.1} & \textbf{84.8 $\pm$ 0.1} \\
& & (r=32) &  &  &  \\
& & ViT & 83.7 $\pm$ 0.1 & 83.9 $\pm$ 0.1 & 83.8 $\pm$ 0.1 \\
\midrule
Experiment 3 & 1. Star-Galaxy & \textit{CNN} & \textit{89.3 $\pm$ 0.1} & \textit{90.3 $\pm$ 0.1} & \textit{89.3 $\pm$ 0.1} \\
& & SENet-CNN & \textbf{90.1 $\pm$ 0.1} & \textbf{91.0 $\pm$ 0.1} & \textbf{90.1 $\pm$ 0.1} \\
& & (r=2) &  &  &  \\
& & ViT & 86.4 $\pm$ 0.1 & 88.3 $\pm$ 0.1 & 86.4 $\pm$ 0.1 \\
& 2. Star-Galaxy-Quasar & \textit{CNN} & \textit{69.4 $\pm$ 0.1} & \textit{73.9 $\pm$ 0.1} & \textit{69.4 $\pm$ 0.1} \\
& & SENet-CNN & \textbf{70.5 $\pm$ 0.1} & \textbf{74.3 $\pm$ 0.1} & \textbf{70.5 $\pm$ 0.1} \\
& & (r=32) &  &  &  \\
& & ViT & 67.9 $\pm$ 0.1 & 73.0 $\pm$ 0.1 & 67.9 $\pm$ 0.1 \\
\bottomrule
\end{tabular}
\end{table}

\subsection{Performance Analysis with FITS Images}
\label{subsec:analysis_image}
We initially employed the inception-based CNN model from C23 as our baseline while working with the FITS images alone. We then enhanced this baseline CNN by integrating attention mechanisms to improve its performance. Particularly, we investigated the incorporation of the SENet as detailed in Section \ref{sec:cls_images} into our baseline model after each inception module. Moreover, our study represents the pioneering exploration of source classification using attention-augmented CNNs and ViT architectures in the literature. Preprocessed images with dimensions of 32$\times$32$\times$5 serve as input for all these models, excluding the photometric features.

\begin{table}[htpb]
\caption{Image-based models with the number of trainable parameters. The listed numbers are with three class (star-galaxy-quasar) classification settings.}
\label{tab:params_image}
\begin{tabular}{@{}lr@{}}
\toprule
Model (Architecture Details) & No. of Parameters \\
\midrule
CNN & 25,543,783 \\
SENet-CNN (r = 2) & 26,295,335 (3 $\%$ $\uparrow$) \\
SENet-CNN (r = 8) & 25,730,031 (0.7 $\%$ $\uparrow$) \\
SENet-CNN (r = 16) & 25,635,607 (0.3 $\%$ $\uparrow$) \\
SENet-CNN (r = 32) & 25,587,815 (0.1 $\%$ $\uparrow$) \\
ViT & 807,203 (96.6 $\%$ $\downarrow$) \\
\bottomrule
\end{tabular}
\end{table}

In our investigation of the SENet-enhanced CNN, we explored the `reduction (r)' hyperparameter within the SENet. We experimented with values ranging from 2, 8, 16, and 32, studying the SENet-CNN with each \rthis{`r'} value in separate experiments and selecting the best-performing model for each case. Similarly, while exploring the ViT, we considered various patch sizes such as 4, 8, and 16, with a 4$\times$4 patch size proving the most effective. We implemented Linformer to alleviate the computational burden, using 12 encoder layers, 8 attention heads, and 64 as the dimensions of the flattened patches. Results are presented in Table \ref{tab:results_images} for the baseline CNN model and the proposed attention-enhanced CNN (with corresponding hyperparameter values for the best-performing model) and ViT-based models. Table \ref{tab:results_images} demonstrates that the SENet-augmented CNN outperforms the baseline CNN across all the experiments and classification tasks, albeit with marginal performance gain across all the metrics. Further, the proposed SENet-CNN shows significant gains over the ViT models, specifically for Experiments 2 and 3. The proposed SENet-CNN demonstrate decent performance gain in Experiment 3, while models trained on bright sources are applied to faint sources. Table \ref{tab:params_image} details the computational complexity for all image-based models. The performance gain with the SENet-augmented CNN comes at the expense of an additional 0.1$\%$ to 3$\%$ parameters. The highest performance gain in terms of accuracy of around 1$\%$ is observed in Experiment 3 for both star-galaxy and star-galaxy-quasar classification, and the lowest gain of 0.1 $\%$ is noticed in Experiment 1 compared to the baseline model. Conversely, the ViT model, comprising only 3.4$\%$ of the parameters of the baseline CNN, achieves comparable performance across all settings, with a maximum performance drop of around 2.9$\%$ and 3.7$\%$ observed in Experiment 3 for the star-galaxy-quasar classification, when compared with the baseline CNN and best-performing SENet-CNN respectively. The lowest performance drop of 0.3 $\%$ and 0.4 $\%$ is observed in Experiment 1 for the star-galaxy separation, compared to the baseline CNN and best-performing SENet-CNN. However, the proposed ViT stands out as the most lightweight image-based model for source classification.


\subsection{Performance Analysis with Photometric Features and FITS Images}
FITS images and photometric features are provided as inputs when working with the ensemble models and the proposed MM ViT. Ensemble model (MargNet), comprises a parallel stack of photometric feature-based and image-based models.  In this study, alongside the baseline CNN model for images (proposed in C23), we explored the proposed SENet-augmented CNN and ViT performance for the image-based model while maintaining the ANN fixed. These pre-trained models were then integrated in parallel through stacking to construct the final network, MargNet. We meticulously fine-tuned the CNN and ANN models for their respective inputs, obviating the need to retrain MargNet. Each class's objects, comprising images and photometric data, are provided as inputs for the CNN and ANN models. The outputs from these models subsequently feed into the ensemble. The outputs from both components are merged, and a statistical combination of these outputs yields the final result. MargNet's architecture at concatenation encompasses 103 neurons, and the corresponding trainable parameters with different image-based models are detailed in Table \ref{tab:params_image_ens}.

While working with the proposed MM ViT model, we incorporated photometric features and FITS images into the same architecture. We used the same ViT architecture studied for image-based classification to achieve this, as detailed in Section \ref{sec:cls_images}. However, a key distinction in using the MM ViT lies in the derivation of the CLS token. Instead of relying solely on image data, we generated the CLS token for the ViT by incorporating photometric features, as depicted in Figure \ref{fig:mmvit}. The parametric specifications for the ViT remain consistent with those discussed in Section \ref{subsec:analysis_image}. Notably, we adjusted the dimension of the final dense layer, with \textit{d} neurons, to 64, matching the dimension of the flattened image patches. Table \ref{tab:params_image_ens} illustrates the computational complexity of the proposed MM ViT. Hybrid models that take both photometric features and images (MargNet-based and MM ViT from Table \ref{tab:results_images_pf}) as input perform better than individual ANN (for photometric features, from Table \ref{tab:results_pf}) and CNN or ViT (for images, from Table \ref{tab:results_images}) models for both star-galaxy and star-galaxy-quasar classifications. The performance of hybrid models across all the experiments described in Section \ref{sec:faint_copmact_sec} are discussed in detail in the sections below.

\begin{table}[h]
\caption{Ensemble models with the number of trainable parameters. The listed numbers are with three class (star-galaxy-quasar) classification settings.}
\label{tab:params_image_ens}
\begin{tabular}{@{}lr@{}}
\toprule
Model & No. of Parameters \\
\midrule
MargNet & 25,875,217 \\
SENet MargNet (r = 2) & 26,626,769 (3 $\%$ $\uparrow$) \\
SENet MargNet (r = 8) & 26,061,465 (0.7 $\%$ $\uparrow$) \\
SENet MargNet (r = 16) & 25,967,041 (0.3 $\%$ $\uparrow$) \\
SENet MargNet (r = 32) & 25,919,249 (0.1 $\%$ $\uparrow$) \\
ViT MargNet & 1,138,637 (96.6 $\%$ $\downarrow$) \\
MM ViT & 1,505,315 (94.2 $\%$ $\downarrow$) \\
\bottomrule
\end{tabular}
\end{table}

\begin{table}[h]
\caption{Quantitative performance of the MargNet and MM ViT models using photometric features and FITS images. While using MargNet, the pre-trained photometric features model (ANN) is fixed and uses different image-based models, including CNN, SENet-augmented CNN, and ViT.}
\label{tab:results_images_pf}
\begin{tabular}{@{}lcccccc@{}}
\toprule
Experiment & Classes & Model & Accuracy ($\%$) & Precision ($\%$) & Recall ($\%$) \\
\midrule
Experiment 1 & 1. Star-Galaxy & \textit{MargNet} & \textit{98.1 $\pm$ 0.1} & \textit{98.1 $\pm$ 0.1} & \textit{98.1 $\pm$ 0.1} \\
& & SENet MargNet & \textbf{98.2 $\pm$ 0.1} & \textbf{98.2 $\pm$ 0.1} & \textbf{98.2 $\pm$ 0.1} \\
& & ViT MargNet & 98.1 $\pm$ 0.1 & 98.1 $\pm$ 0.1 & 98.1 $\pm$ 0.1 \\
& & MM ViT & 98.1 $\pm$ 0.1 & 98.1 $\pm$ 0.1 & 98.1 $\pm$ 0.1 \\
& 2. Star-Galaxy-Quasar & \textit{MargNet} & \textit{93.3 $\pm$ 0.2} & \textit{93.3 $\pm$ 0.2} & \textit{93.3 $\pm$ 0.2} \\
& & SENet MargNet & \textbf{93.5 $\pm$ 0.2} & \textbf{93.5 $\pm$ 0.2} & \textbf{93.5 $\pm$ 0.2} \\
& & ViT MargNet & 93.2 $\pm$ 0.2 & 93.2 $\pm$ 0.2 & 93.2 $\pm$ 0.2 \\
& & MM ViT & 93.2 $\pm$ 0.2 & 93.2 $\pm$ 0.2 & 93.2 $\pm$ 0.2 \\
\midrule
Experiment 2 & 1. Star-Galaxy & \textit{MargNet} & \textit{96.9 $\pm$ 0.1} & \textit{96.9 $\pm$ 0.1} & \textit{96.9 $\pm$ 0.1} \\
& & SENet MargNet & \textbf{97.1 $\pm$ 0.1} & \textbf{97.1 $\pm$ 0.1} & \textbf{97.1 $\pm$ 0.1} \\
& & ViT MargNet & 96.8 $\pm$ 0.1 & 96.8 $\pm$ 0.1 & 96.8 $\pm$ 0.1 \\
& & MM ViT & 96.9 $\pm$ 0.1 & 96.9 $\pm$ 0.1 & 96.9 $\pm$ 0.1 \\
& 2. Star-Galaxy-Quasar & \textit{MargNet} & \textit{87.3 $\pm$ 0.2} & \textit{87.4 $\pm$ 0.2} & \textit{87.3 $\pm$ 0.2} \\
& & SENet MargNet & \textbf{87.5 $\pm$ 0.2} & \textbf{87.5 $\pm$ 0.2} & \textbf{87.5 $\pm$ 0.2} \\
& & ViT MargNet & 86.9 $\pm$ 0.2 & 86.9  $\pm$ 0.2 & 86.9 $\pm$ 0.2 \\
& & MM ViT & 86.3 $\pm$ 0.2 & 86.2 $\pm$ 0.2 & 86.3 $\pm$ 0.2 \\
\midrule
Experiment 3 & 1. Star-Galaxy & \textit{MargNet} & \textit{92.0 $\pm$ 0.1} & \textit{92.8 $\pm$ 0.1} & \textit{92.0 $\pm$ 0.1} \\
& & SENet MargNet & \textbf{92.9 $\pm$ 0.1} & \textbf{93.4 $\pm$ 0.1} & \textbf{92.9 $\pm$ 0.1} \\
& & ViT MargNet & 91.7 $\pm$ 0.1 & 92.6 $\pm$ 0.1 & 91.7 $\pm$ 0.1 \\
& & MM ViT & 91.8 $\pm$ 0.1 & 92.5 $\pm$ 0.1 & 91.8 $\pm$ 0.1 \\
& 2. Star-Galaxy-Quasar & \textit{MargNet} & \textit{73.1 $\pm$ 0.2} & \textit{76.4 $\pm$ 0.2} & \textit{73.1 $\pm$ 0.2} \\
& & SENet MargNet & \textbf{73.2 $\pm$ 0.2} & \textbf{76.5 $\pm$ 0.2} & \textbf{73.2 $\pm$ 0.2} \\
& & ViT MargNet & 72.7 $\pm$ 0.1 & 76.3 $\pm$ 0.1 & 72.7 $\pm$ 0.2 \\
& & MM ViT & 71.8 $\pm$ 0.2 & 75.4 $\pm$ 0.2 & 71.8 $\pm$ 0.2 \\
\bottomrule
\end{tabular}
\end{table}

\begin{figure}[h]
    \centering
    \begin{subfigure}[b]{0.48\linewidth}
        \centering
        \includegraphics[width=\linewidth]{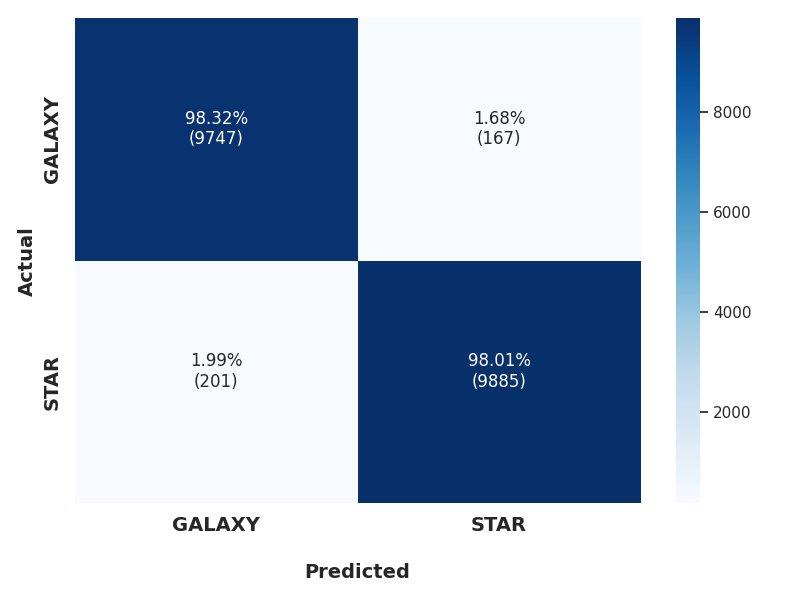}
        \caption{MargNet.}
        \label{fig:ex1_sg_mg}
    \end{subfigure}
    \hfill
    \begin{subfigure}[b]{0.48\linewidth}
        \centering
        \includegraphics[width=\linewidth]{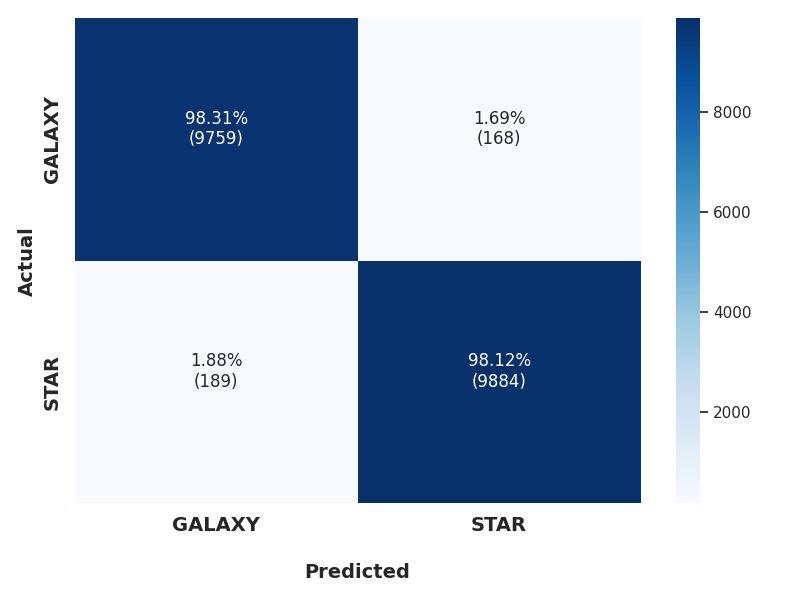}
        \caption{SENet-MargNet.}
        \label{fig:ex1_sg_attmg}
    \end{subfigure}
    
    \vskip\baselineskip
    
    \begin{subfigure}[b]{0.48\linewidth}
        \centering
        \includegraphics[width=\linewidth]{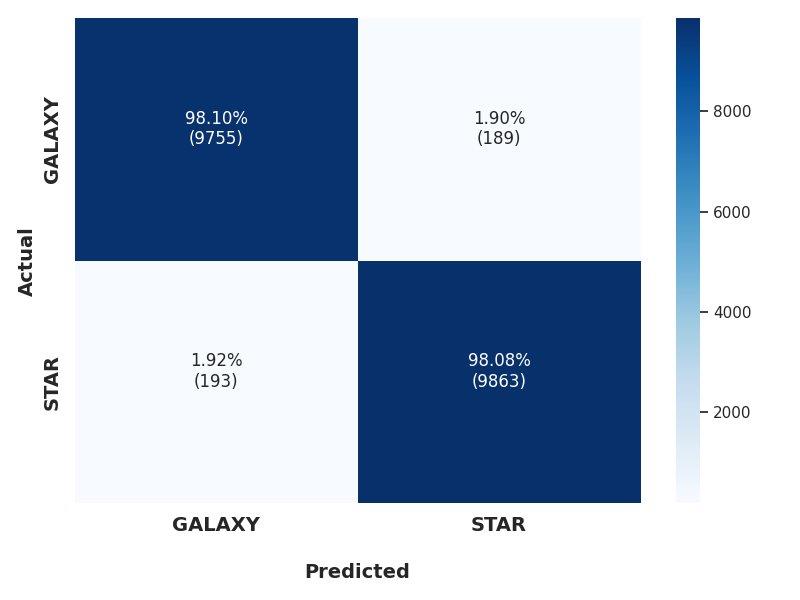}
        \caption{ViT-MargNet.}
        \label{fig:ex1_sg_vitmg}
    \end{subfigure}    
    \hfill
    \begin{subfigure}[b]{0.48\linewidth}
        \centering
        \includegraphics[width=\linewidth]{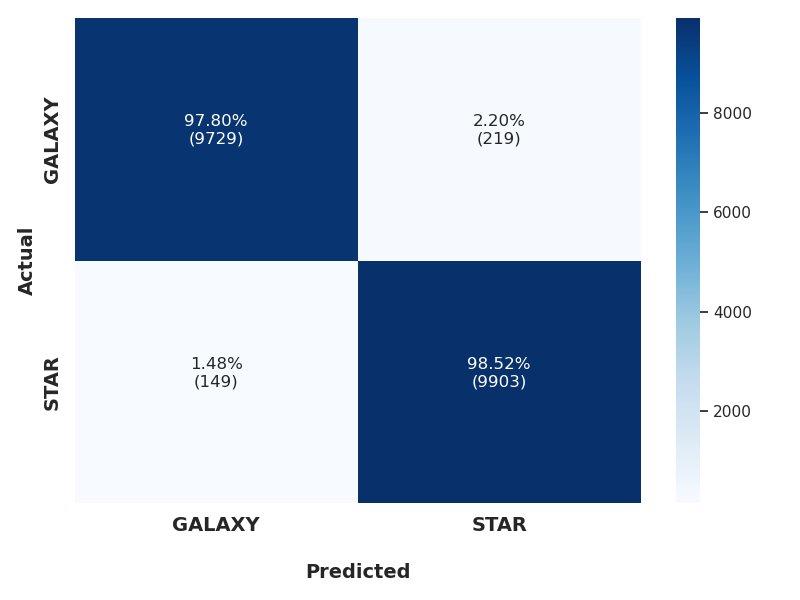}
        \caption{MM ViT.}
        \label{fig:ex1_sg_mmvitmg}
    \end{subfigure}
    
    \caption{Confusion matrices of MargNet (a), SENet-augmented MargNet (b), ViT-MargNet (c), and MM ViT (right) for Experiment 1, depicting star-galaxy classification performance. The proposed SENet-MargNet outperforms all other models and achieves 98.31$\%$ accuracy for galaxies and 98.12$\%$ for stars. Minimal misclassifications among stars and galaxies are observed, with rates as low as approximately 1.78$\%$, 1.83$\%$, 1.91$\%$, and 1.84$\%$ for SENet-MargNet, MargNet, ViT-MargNet, and MM ViT, respectively.}
    \label{fig:ex1_sg_cm}
\end{figure}

\begin{figure}[h]
    \centering
    \begin{subfigure}[b]{0.48\linewidth}
        \centering
        \includegraphics[width=\linewidth]{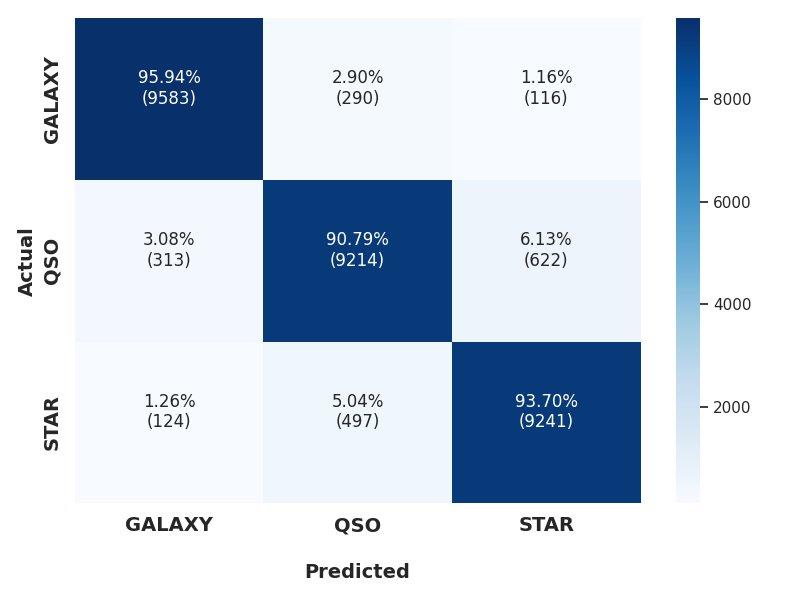}
        \caption{MargNet.}
        \label{fig:ex1_sgq_mg}
    \end{subfigure}
    \hfill
    \begin{subfigure}[b]{0.48\linewidth}
        \centering
        \includegraphics[width=\linewidth]{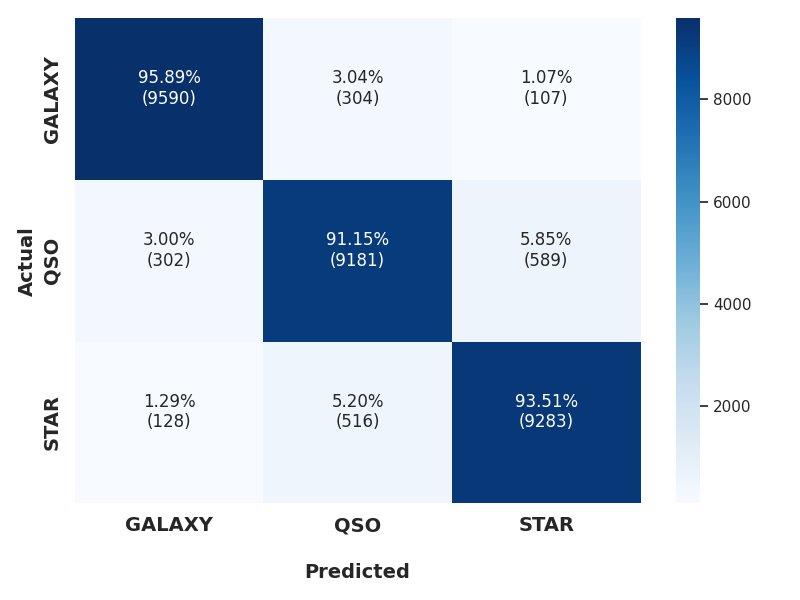}
        \caption{SENet-MargNet.}
        \label{fig:ex1_sgq_attmg}
    \end{subfigure}
    
    \vskip\baselineskip
    
    \begin{subfigure}[b]{0.48\linewidth}
        \centering
        \includegraphics[width=\linewidth]{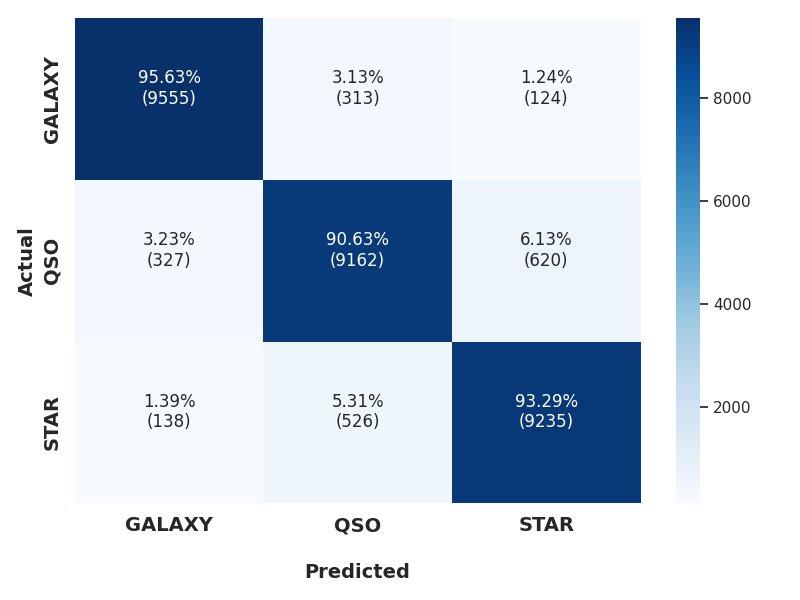}
        \caption{ViT-MargNet.}
        \label{fig:ex1_sgq_vitmg}
    \end{subfigure}    
    \hfill
    \begin{subfigure}[b]{0.48\linewidth}
        \centering
        \includegraphics[width=\linewidth]{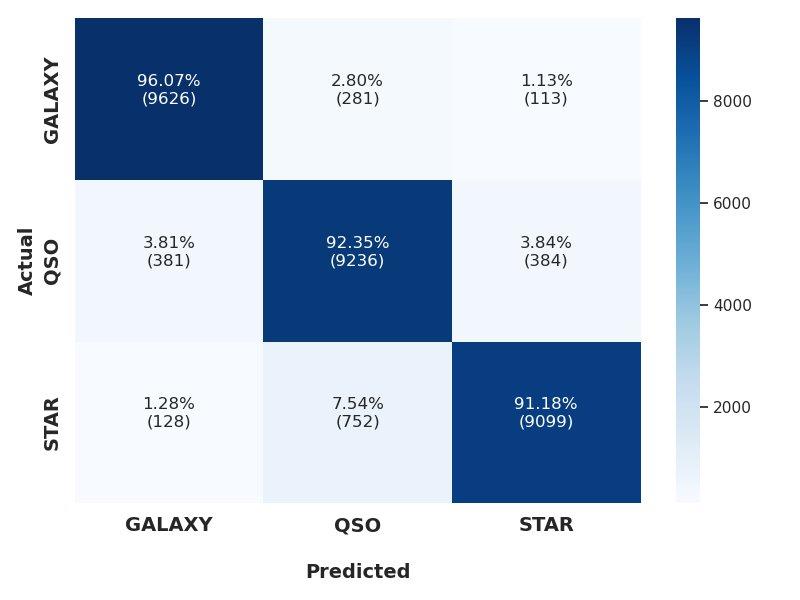}
        \caption{MM ViT.}
        \label{fig:ex1_sgq_mmvitmg}
    \end{subfigure}
    
    \caption{Confusion matrices are presented for MargNet (a), SENet-augmented MargNet (b), ViT-MargNet (c), and MM ViT (right) in Experiment 1, illustrating the classification performance for star-galaxy-quasar categorization. The proposed SENet-MargNet performs better than all other models, achieving 95.89$\%$ accuracy for galaxies, 91.15$\%$ for quasars, and 93.51$\%$ for stars. Notably, minimal misclassifications are observed among stars, quasars, and galaxies, with rates as low as approximately 3.24$\%$, 3.26$\%$, 3.41$\%$, and 3.4$\%$ for SENet-MargNet, MargNet, ViT-MargNet, and MM ViT, respectively.}
    \label{fig:ex1_sgq_cm}
\end{figure}

\subsubsection{Results: Experiment 1}
This experiment selects all the three sets: training, validation, and testing from the Compact Source dataset. Performance metrics are computed for star-galaxy and star-galaxy-quasar classifications as illustrated in Table \ref{tab:results_images_pf}. For star-galaxy classification, quasars are excluded to facilitate comparison with only star-galaxy separation models used in other studies. The proposed SENet-augmented MargNet achieves an overall accuracy of $98.2 \pm 0.1\%$, while MargNet, ViT-MargNet, and MM ViT each achieve an average accuracy of $98.1 \pm  0.1\%$. We can also see from Figure \ref{fig:ex1_sg_cm} that the performance on both stars and galaxies is similar (98.31$\%$ for galaxies and 98.12$\%$ for stars) and higher with SENet-MargNet. MargNet exhibits a slightly higher accuracy for galaxies at 93.32$\%$ while maintaining an accuracy of 98.01$\%$ for stars. Meanwhile, ViT-MargNet and MM ViT demonstrate performance with an accuracy of 98.1$\%$ and 97.8$\%$ for galaxies 98.08$\%$ and 98.52$\%$ for stars, respectively. Notably, misclassifications between stars and galaxies are minimal, with MargNet, SENet-MargNet, ViT-MargNet and MM ViT achieving rates as low as approximately 1.83$\%$, 1.78$\%$, 1.91$\%$ and 1.84$\%$, respectively.

For the task of star-galaxy-quasar classification, the proposed SENet-MargNet achieves an average accuracy of $93.5\pm0.2\%$, with similar accuracies of $93.3\pm0.2\%$, $93.2\pm0.2\%$ and $93.2\pm0.2\%$ observed for the MargNet, ViT-MargNet and MM ViT models, respectively. However, detailed analysis reveals variations in performance across the individual classes. The confusion matrices in Figure \ref{fig:ex1_sgq_cm} highlight distinct accuracies for stars, galaxies, and quasars. Quasars exhibit the lowest accuracy when classified by the ViT-MargNet model, achieving 90.63$\%$ accuracy, while MM ViT achieves the best accuracy of 92.35$\%$.  An accuracy of 90.79$\%$ and 91.15$\%$ are achieved using MargNet and SENet-MargNet, respectively. Notably, the MM ViT model demonstrates the lowest accuracy for stars, with a rate of 91.18$\%$, and ViT-MargNet demonstrates the lowest accuracy for galaxies at 95.63 $\%$. Further analysis reveals specific misclassifications within each model. For SENet-MargNet, quasars are misclassified as stars in approximately 5.85$\%$ of cases and galaxies in around 3$\%$. The MargNet model misidentifies quasars as stars in 6.13$\%$ of cases and galaxies in 3.08$\%$. The ViT-MargNet misclassifies quasars as stars at 6.13$\%$ and galaxies at 3.23$\%$ of rates. MM ViT misclassifies quasars as stars and galaxies at the same rates of 3.84$\%$ and 3.81$\%$, respectively. The more frequent identification of quasars as stars is understandable, as both are point sources. For the same reason, stars are misclassified as quasars in $\sim5.04\%$, 5.20$\%$, 5.31$\%$ and 7.54$\%$ of the cases for the MargNet, SENet-MargNet, ViT-MargNet and MM ViT, respectively. Galaxies have the best individual accuracy among the three classes and across all the models. Galaxy-quasar misclassifications (galaxy-quasar: 2.90$\%$ and quasar-galaxy: 3.08$\%$ for MargNet, galaxy-quasar: 3.04$\%$ and quasar-galaxy: 3.0$\%$ for SENet-MargNet, galaxy-quasar: 3.13$\%$ and quasar-galaxy: 3.23$\%$ for ViT-MargNet and galaxy-quasar: 2.80$\%$ and quasar-galaxy: 3.81$\%$ for MM ViT being the individual misclassification rates) are more common than galaxy-star misclassifications (galaxy-star: 1.16$\%$ and star-galaxy: 1.26$\%$ for MargNet, galaxy-star: 1.07$\%$ and star-galaxy: 1.29$\%$ for SENet-MargNet, galaxy-star: 1.24$\%$ and star-galaxy: 1.39$\%$ for ViT-MargNet and galaxy-star: 1.13$\%$ and star-galaxy: 1.28$\%$ for MM ViT). This arises from the quasar host galaxy fuzz~\citep{jahnke2003b}, which may be observable particularly with low-luminosity quasars, sometimes resulting in misclassification as a compact galaxy. Stars, devoid of structure beyond the Point Spread Function (PSF), are generally less susceptible to being misclassified as galaxies.

\begin{figure*}[!htpb]
    \centering
    \begin{subfigure}[b]{0.48\linewidth}
        \centering
        \includegraphics[width=\linewidth]{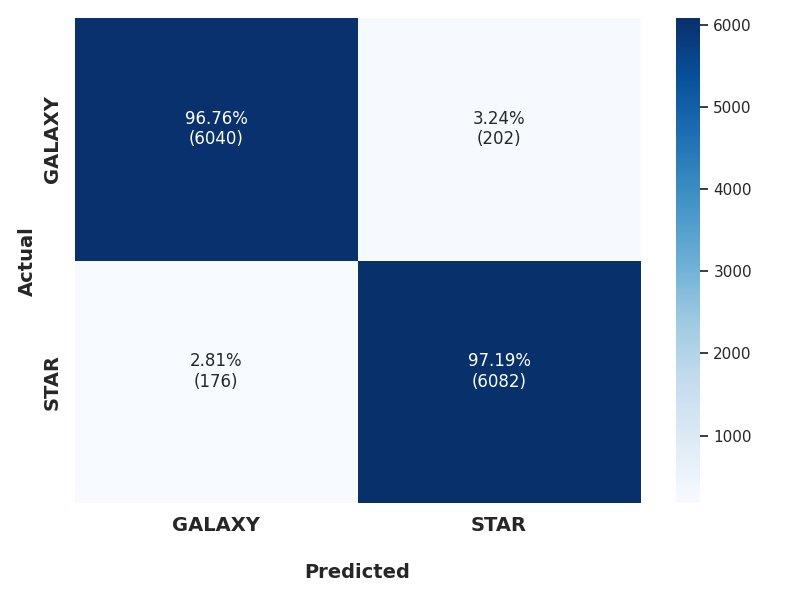}
        \caption{MargNet.}
        \label{fig:ex2_sg_mg}
    \end{subfigure}
    \hfill
    \begin{subfigure}[b]{0.48\linewidth}
        \centering
        \includegraphics[width=\linewidth]{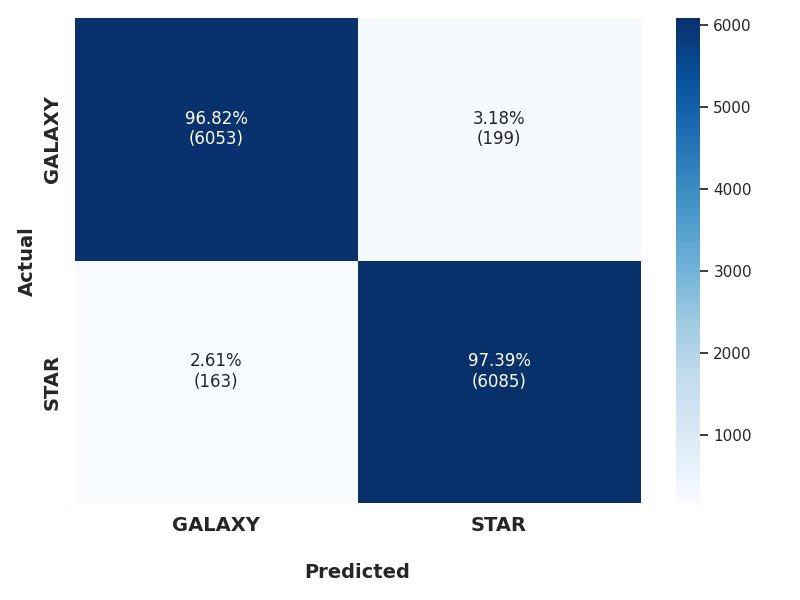}
        \caption{SENet-MargNet.}
        \label{fig:ex2_sg_attmg}
    \end{subfigure}
    
    \vskip\baselineskip
    
    \begin{subfigure}[b]{0.48\linewidth}
        \centering
        \includegraphics[width=\linewidth]{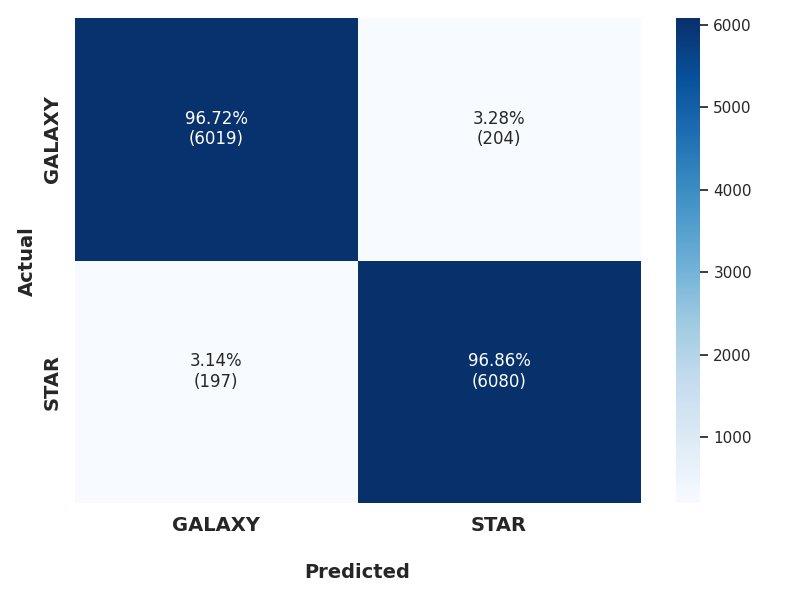}
        \caption{ViT-MargNet.}
        \label{fig:ex2_sg_vitmg}
    \end{subfigure}    
    \hfill
    \begin{subfigure}[b]{0.48\linewidth}
        \centering
        \includegraphics[width=\linewidth]{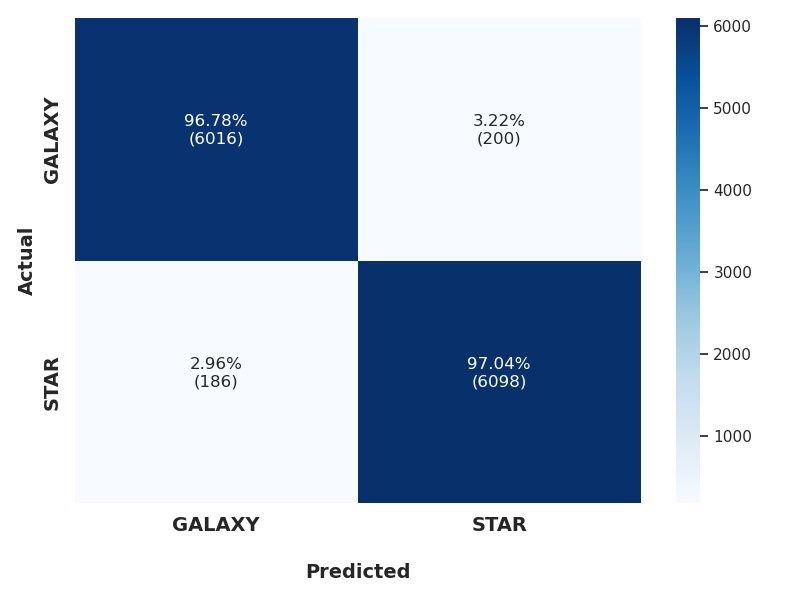}
        \caption{MM ViT.}
        \label{fig:ex2_sg_mmvitmg}
    \end{subfigure}
    
    \caption{Confusion matrices are provided for MargNet (a), SENet-augmented MargNet (b), ViT-MargNet (c), and MM ViT (d) in Experiment 2, focusing on star-galaxy classification performance. SENet-MargNet achieves high accuracy rates of 96.82$\%$ for classifying galaxies and 97.39$\%$ for stars. Importantly, misclassifications between stars and galaxies are minimal across all models, with SENet-MargNet, MargNet, ViT-MargNet, and MM ViT achieving rates as low as approximately 2.89$\%$, 3.02$\%$, 3.21$\%$, and 3.09$\%$, respectively.}
    \label{fig:ex2_sg_cm}
\end{figure*}

\begin{figure*}[!htpb]
    \centering
    \begin{subfigure}[b]{0.48\linewidth}
        \centering
        \includegraphics[width=\linewidth]{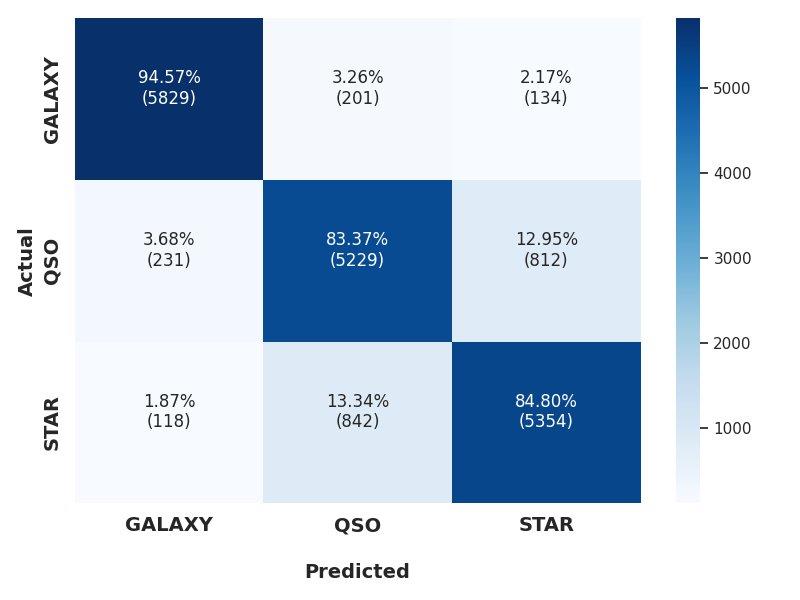}
        \caption{MargNet.}
        \label{fig:ex2_sgq_mg}
    \end{subfigure}
    \hfill
    \begin{subfigure}[b]{0.48\linewidth}
        \centering
        \includegraphics[width=\linewidth]{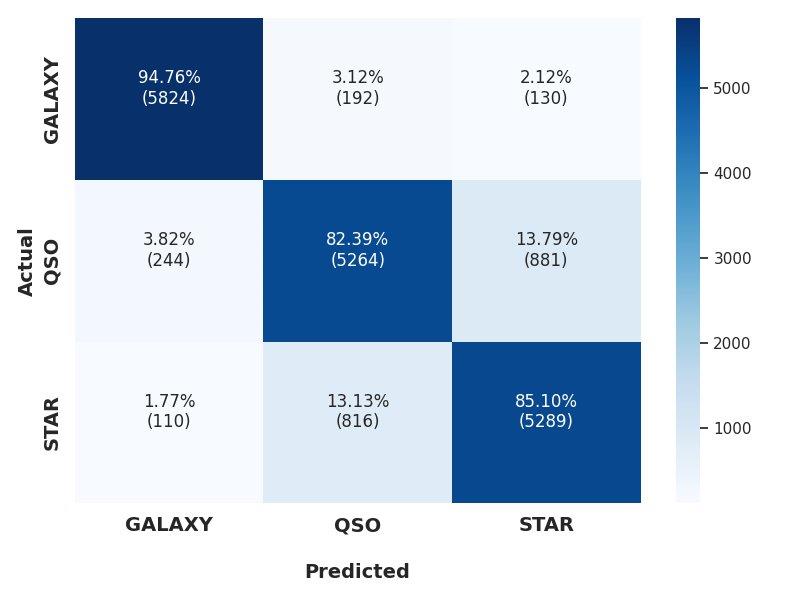}
        \caption{SENet-MargNet.}
        \label{fig:ex2_sgq_attmg}
    \end{subfigure}
    
    \vskip\baselineskip
    
    \begin{subfigure}[b]{0.48\linewidth}
        \centering
        \includegraphics[width=\linewidth]{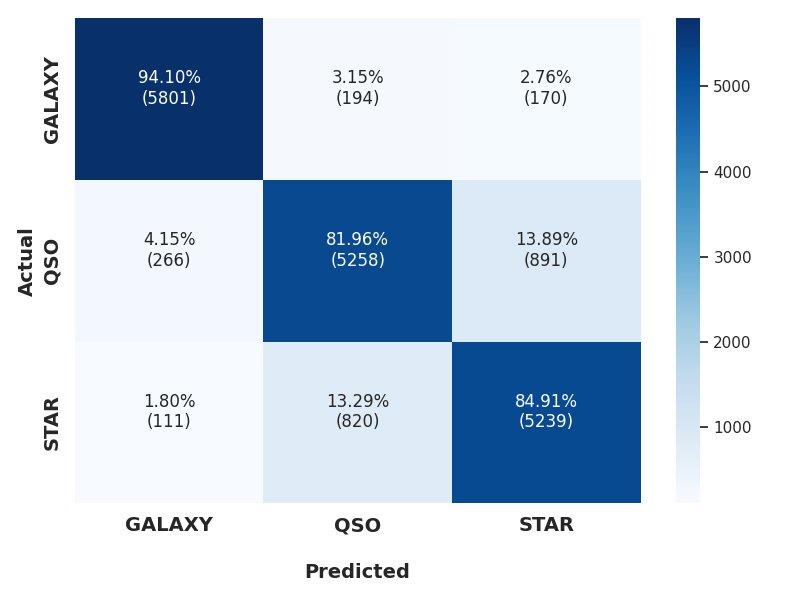}
        \caption{ViT-MargNet.}
        \label{fig:ex2_sgq_vitmg}
    \end{subfigure}    
    \hfill
    \begin{subfigure}[b]{0.48\linewidth}
        \centering
        \includegraphics[width=\linewidth]{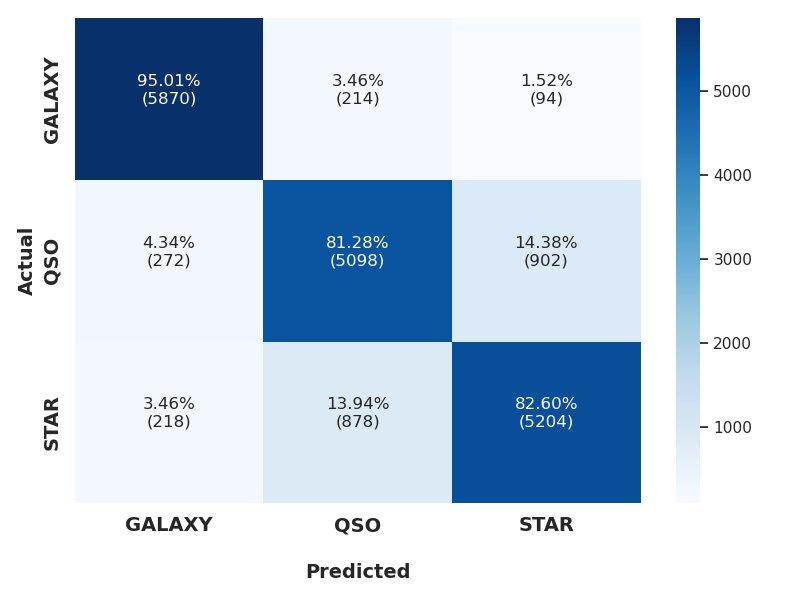}
        \caption{MM ViT.}
        \label{fig:ex2_sgq_mmvitmg}
    \end{subfigure}
    
    \caption{Confusion matrices comparing MargNet (a), SENet-augmented MargNet (b), ViT-MargNet (c) and MM ViT (d) for Experiment 2 reveal their performance in star-galaxy-quasar classification. SENet-MargNet achieves 94.57$\%$ accuracy for galaxies, 83.37$\%$ for quasars, and 84.8$\%$ for stars. Notably, misclassifications among stars, quasars, and galaxies are minimal, with SENet-MargNet, MargNet, and MM ViT achieving error rates as low as approximately 6.21$\%$, 6.29$\%$, 6.5$\%$ and 6.85$\%$, respectively.}
    \label{fig:ex2_sgq_cm}
\end{figure*}

\subsubsection{Results: Experiment 2}
\begin{figure*}[htpb]
    \centering
    \begin{subfigure}[b]{0.465\linewidth}
        \centering
        \includegraphics[width=\linewidth]{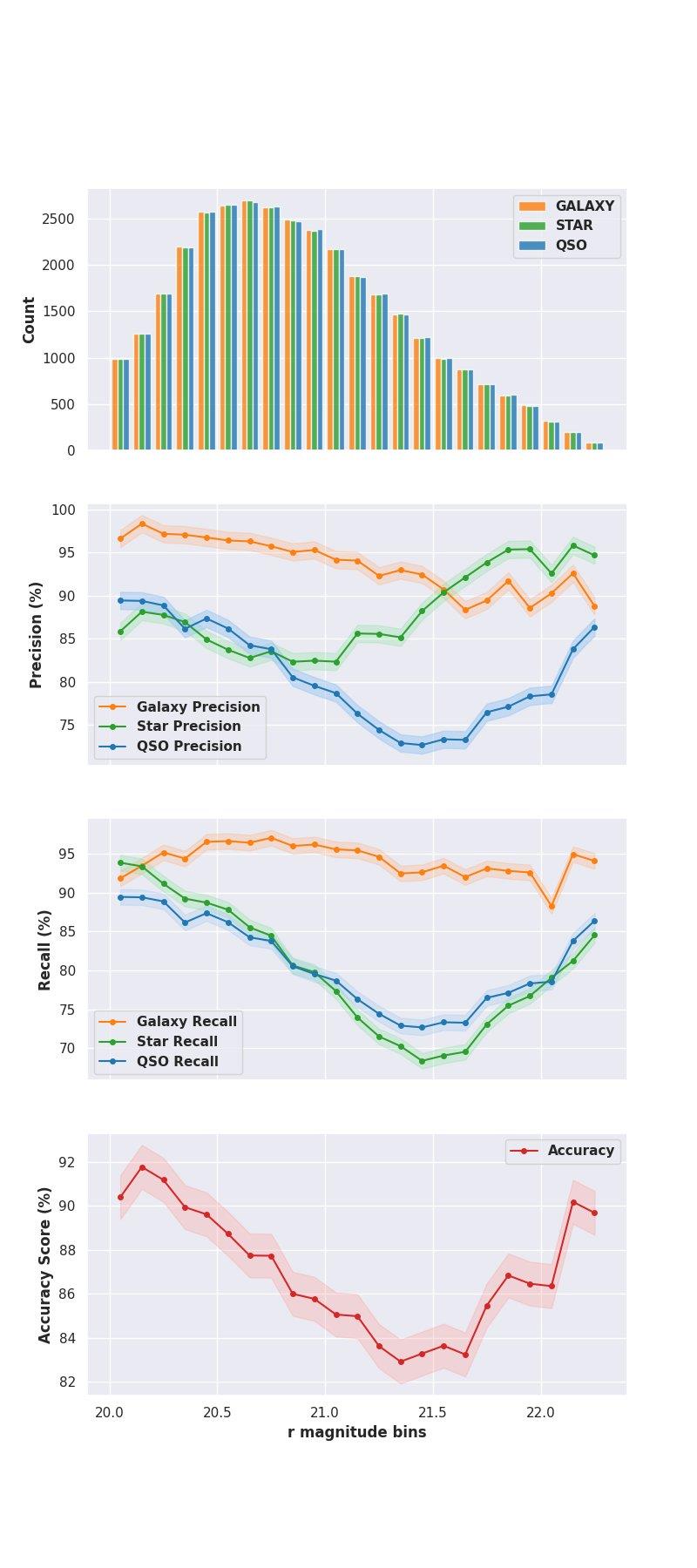}
        \caption{star–galaxy-quasar (SGQ) classification performance using MargNet for Experiment 2: We see accuracy sharply decreases till \textit{r} = 21 and an unexpected rise for \textit{r} $>$ 22. For star recall, quasar recall,  and quasar precision, a rise follows a drop can be seen. At \textit{r} $>$ 22, there is no rise seen in galaxy precision.}
        \label{fig:ex2_sgq_mg1}
    \end{subfigure}
    \hfill
    \begin{subfigure}[b]{0.465\linewidth}
        \centering
        \includegraphics[width=\linewidth]{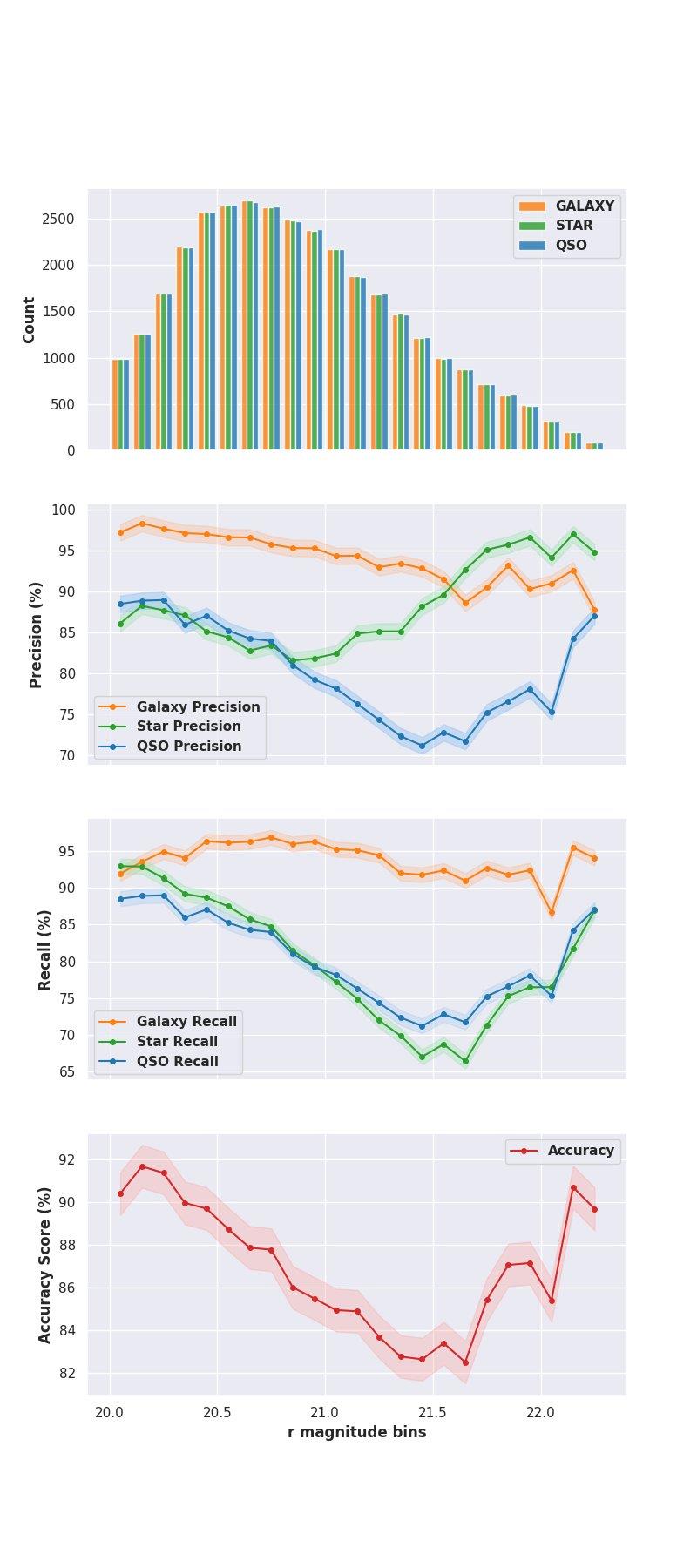}
        \caption{star–galaxy-quasar (SGQ) classification performance using SENet-MargNet for Experiment 2: We see accuracy sharply decreases till \textit{r} = 21 and an unexpected rise for \textit{r} $>$ 22. For star recall, quasar recall,  and quasar precision, a rise follows a drop can be seen. At \textit{r} $>$ 22, there is no rise seen in galaxy precision.}
        \label{fig:ex2_sgq_attmg1}
    \end{subfigure}
    \caption{In Experiment 2, performance on the test data is assessed for the faint and compact dataset (\textit{C} $<$ 0.5; \textit{r} $>$ 20), segmented into bins of 0.1 magnitude width, with the final range being 20 $<$ \textit{r} $<$ 22.6. Precision, recall, and accuracy metrics are computed for the star-galaxy-quasar classification problem, as depicted in Figure \ref{fig:ex2_sgq_mg1} and Figure \ref{fig:ex2_sgq_attmg1} for the MargNet and SENet-MargNet models, respectively. From the accuracy plot, it is observed that the proposed SENet-MargNet is performing slightly better than the baseline MargNet model.}
    \label{fig:ex2_sgq_mg_attmg1}
\end{figure*}

\begin{figure*}[htpb]
    \centering
    \begin{subfigure}[b]{0.47\linewidth}
        \centering
        \includegraphics[width=\linewidth]{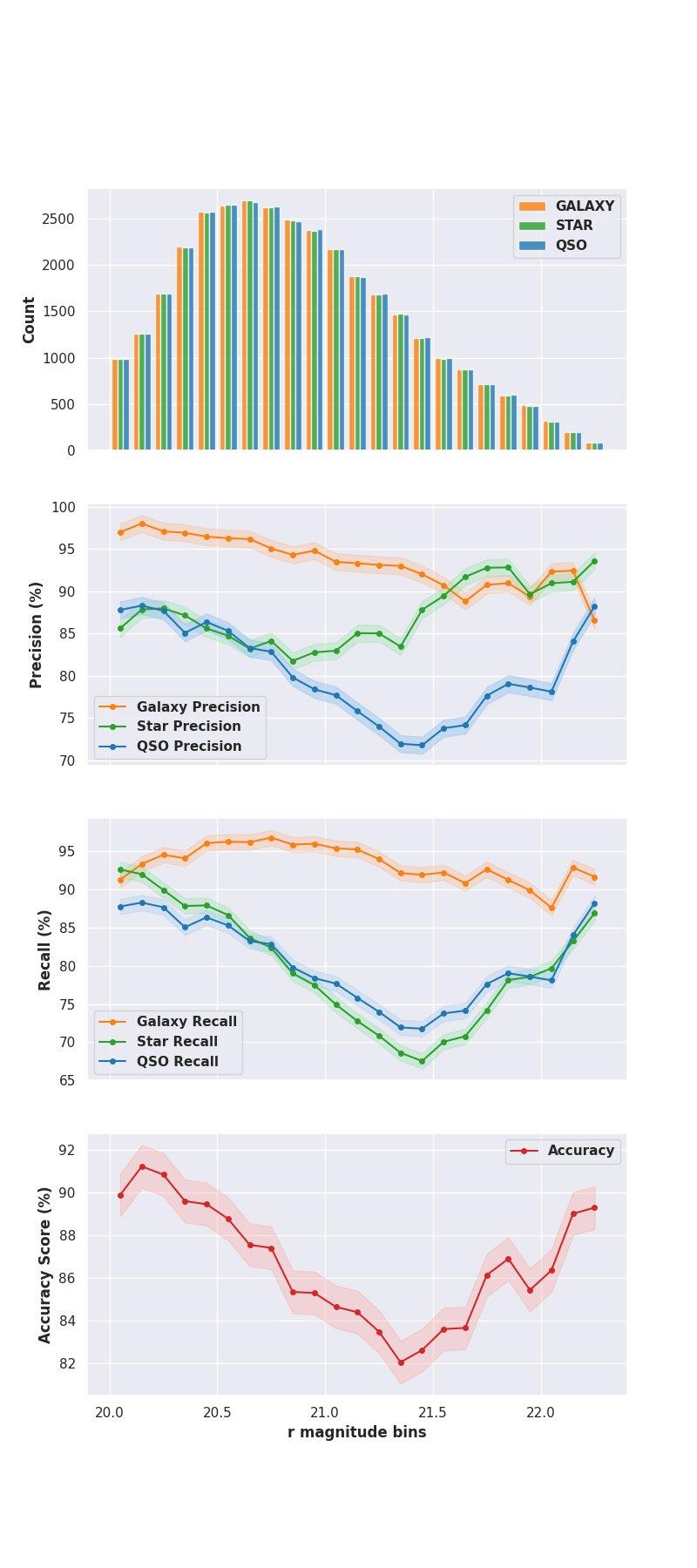}
        \caption{star–galaxy-quasar (SGQ) classification performance using ViT-MargNet for Experiment 2: We see accuracy sharply decreases till \textit{r} = 21 and an unexpected rise for \textit{r} $>$ 22. For star recall, quasar recall,  and quasar precision, a rise follows a drop can be seen. At \textit{r} $>$ 22, there is no rise seen in galaxy precision.}
        \label{fig:ex2_sgq_vit1}
    \end{subfigure}
    \hfill
    \begin{subfigure}[b]{0.47\linewidth}
        \centering
        \includegraphics[width=\linewidth]{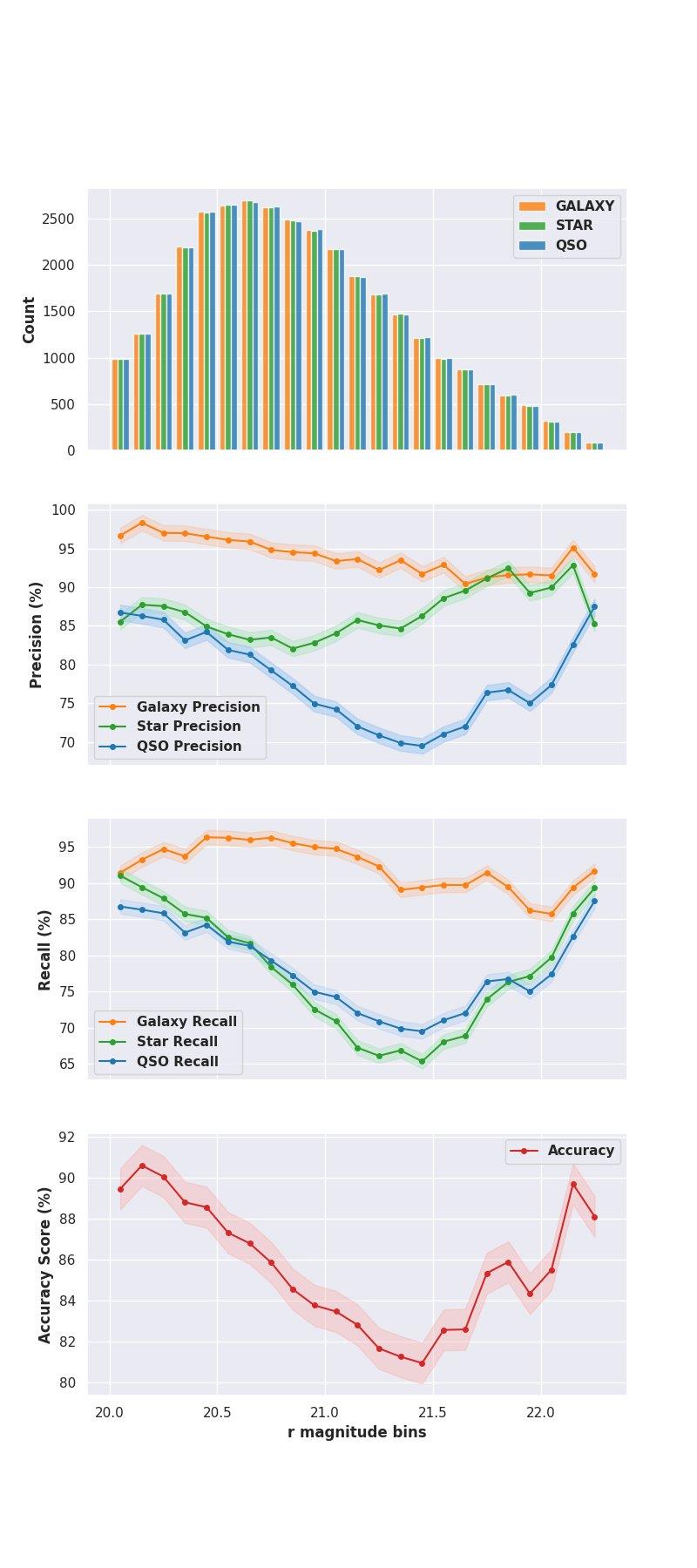}
        \caption{star–galaxy-quasar (SGQ) classification performance using MM ViT for Experiment 2: We see accuracy sharply decreases till \textit{r} = 21 and an unexpected rise for \textit{r} $>$ 22. For star recall, quasar recall,  and quasar precision, a rise follows a drop can be seen. At \textit{r} $>$ 22, there is no rise seen in galaxy precision.}
        \label{fig:ex2_sgq_mmvit1}
    \end{subfigure}
    \caption{In Experiment 2, performance on the test data is assessed for the faint and compact dataset (\textit{C} $<$ 0.5; \textit{r} $>$ 20), segmented into bins of 0.1 magnitude width, with the final range being 20 $<$ \textit{r} $<$ 22.6. Precision, recall, and accuracy metrics are computed for the star-galaxy-quasar classification problem, as depicted in Figure \ref{fig:ex2_sgq_vit1} and Figure \ref{fig:ex2_sgq_mmvit1} for the ViT-MargNet and MM ViT models, respectively. The ViT-MargNet and MM ViT models perform very similarly to the baseline MargNet models.}
    \label{fig:ex2_sgq_vit_mmvit1}
\end{figure*}

In Experiment 2, all datasets; training, validation, and test—are derived from the Faint and Compact Source dataset. Similar to Experiment 1, various models are trained and evaluated using the Experiment 2 dataset, and their performance metrics are summarized in Table \ref{tab:results_images_pf}. Across all metrics, there is a decline in performance compared to Experiment 1, attributed to the lower signal-to-noise ratio (SNR) as objects become fainter. This noise affects photometric measurements and feature calculations, posing challenges in classification. For star-galaxy separation, SENet-MargNet achieves 97.1$\pm$0.1$\%$ accuracy, while MargNet, ViT-MargNet, and MM ViT achieve 96.9$\pm$0.1$\%$, 96.8$\pm$0.1$\%$, and 96.9$\pm$0.1$\%$, respectively. Confusion matrices depicted in Figure \ref{fig:ex2_sg_cm} reveal SENet-MargNet's excellence with 96.82$\%$ accuracy for galaxies and 97.39$\%$ for stars. MargNet shows 96.76$\%$ accuracy for galaxies and 97.19$\%$ for stars. ViT-MargNet and MM ViT score 96.72$\%$ and 96.78$\%$ for galaxies and 96.86$\%$ and 97.04$\%$ for stars, respectively. Minimal misclassifications between stars and galaxies are observed, with rates of approximately 2.89$\%$, 3.02$\%$, 3.21$\%$, and 3.09$\%$ for SENet-MargNet, MargNet, ViT-MargNet, and MM ViT, respectively.

SENet-MargNet achieves an average accuracy of $87.5 \pm  0.2$\% for star-galaxy-quasar classification, similar to MargNet, ViT-MargNet, and MM ViT, which achieve accuracies of $87.3\pm0.2$\%, 86.9$\pm$0.2$\%$, and 86.3$\pm$0.2$\%$, respectively. Individual class accuracies are depicted in Figure \ref{fig:ex2_sgq_cm}. Quasars display the lowest accuracy with MM ViT at 81.28$\%$ and the highest with SENet-MargNet at 83.37$\%$. Stars achieve the lowest accuracy with MM ViT at 82.6$\%$ and the highest with MargNet at 85.1$\%$. Similarly, for galaxies, the lowest accuracy of 94.1$\%$ is achieved using ViT-MargNet and the highest with MM ViT at 95.01$\%$. Notably, misclassifications among stars, quasars, and galaxies are minimal, with SENet-MargNet, MargNet, and MM ViT achieving error rates as low as approximately 6.21$\%$, 6.29$\%$, 6.5$\%$, and 6.85$\%$, respectively. We further evaluate the performance statistics as a function of magnitude to assess model behaviour at fainter levels. The test set is divided into bins of 0.1 magnitudes within the $20 < r < 22.6$ range, ensuring each bin contains at least 50 objects per class for robust analysis. Metrics are assessed for each bin and visualized in Figure \ref{fig:ex2_sgq_mg_attmg1} and Figure \ref{fig:ex2_sgq_vit_mmvit1}. For star-galaxy-quasar classification, Figure \ref{fig:ex2_sgq_mg1} (MargNet model) shows accuracy steadily decreasing up to $r = 21.3$, in line with diminishing SNR. 

\rthis{Across all the models, the accuracy starts increasing beyond $r > 21.3$ for both star-galaxy and star-galaxy-quasar classifications. This is unexpected, as fainter galaxies are usually harder to classify and thus should lower the accuracy. Similar trends were observed in C23. The discrepancy might stem from the artifact in the training dataset distribution across different $r$ magnitudes, containing 2831 galaxies, 190 stars, and 660 quasars for $r > 21.3$, despite equal class representation overall. On the other hand, the testing dataset maintains an equal class distribution across magnitude bins to avoid bias. This training bias reduces false negatives for galaxies, leading to fewer false positives for stars and quasars, thus potentially boosting the precision at fainter magnitudes.} Similar trends are observed for quasar precision, quasar recall, and star recall, with a rise following a drop. Star precision and galaxy recall remain consistently high. However, galaxy precision behaves differently; at $\textit{r} > 22$, no increase is observed, plateauing. A similar trend is observed for SENet-MargNet (in Figure \ref{fig:ex2_sgq_attmg1}), ViT-Margnet (in Figure \ref{fig:ex2_sgq_vit1}) and MM ViT (in Figure \ref{fig:ex2_sgq_mmvit1}) models.

\subsubsection{Results: Experiment 3}
\begin{figure*}[htpb]
    \centering
    \begin{subfigure}[b]{0.48\linewidth}
        \centering
        \includegraphics[width=\linewidth]{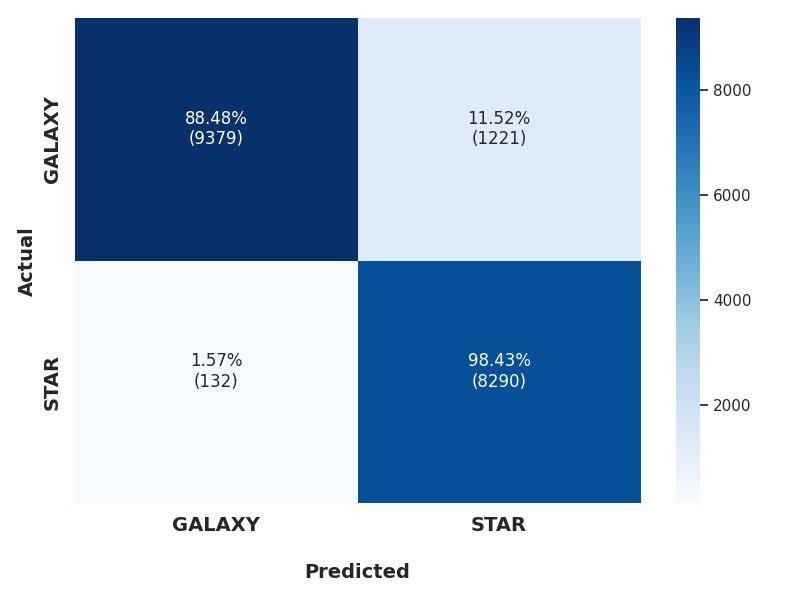}
        \caption{MargNet.}
        \label{fig:ex3_sg_mgn}
    \end{subfigure}
    \hfill
    \begin{subfigure}[b]{0.48\linewidth}
        \centering
        \includegraphics[width=\linewidth]{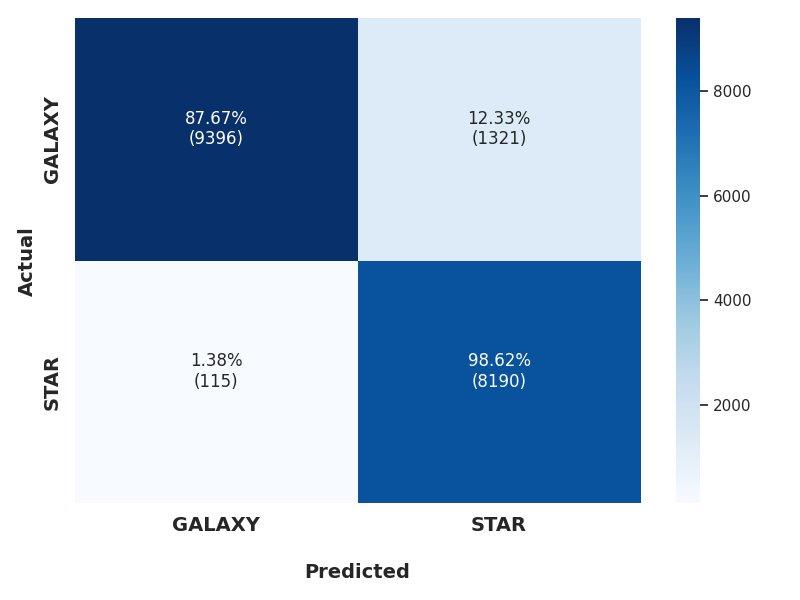}
        \caption{SENet-MargNet.}
        \label{fig:ex3_sg_attmgn}
    \end{subfigure}
    
    \vskip\baselineskip
    
    \begin{subfigure}[b]{0.48\linewidth}
        \centering
        \includegraphics[width=\linewidth]{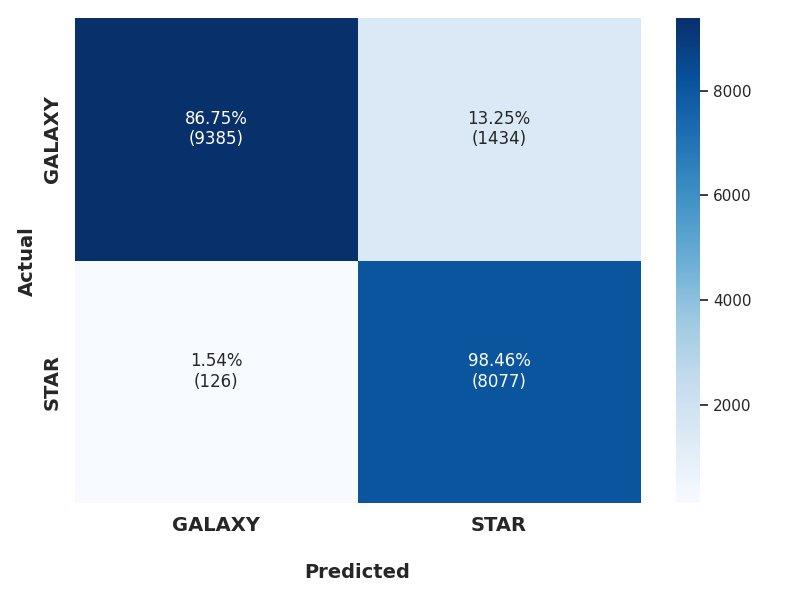}
        \caption{ViT-MargNet.}
        \label{fig:ex3_sg_vitmgn}
    \end{subfigure}    
    \hfill
    \begin{subfigure}[b]{0.48\linewidth}
        \centering
        \includegraphics[width=\linewidth]{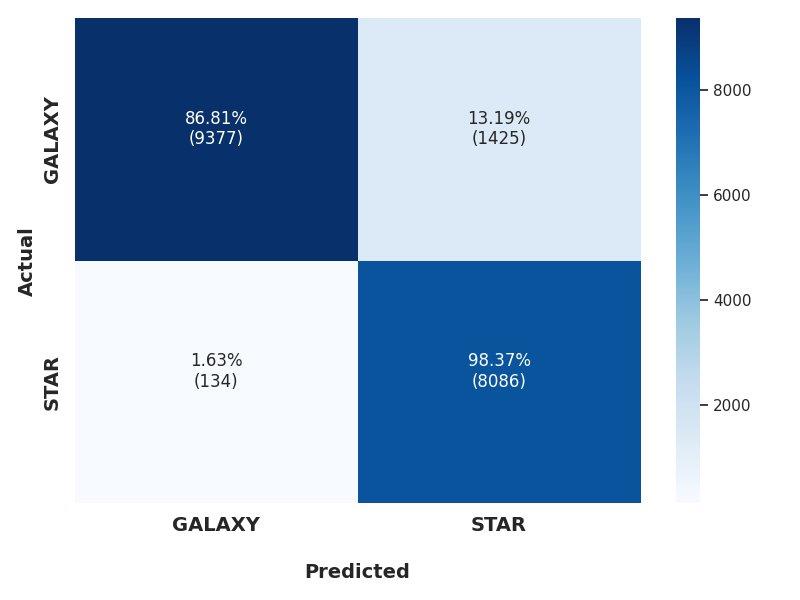}
        \caption{MM ViT.}
        \label{fig:ex3_sg_mmvitmgn}
    \end{subfigure}
    
    \caption{Confusion matrices are provided for MargNet (a), SENet-augmented MargNet (b), ViT-MargNet (c), and MM ViT (d) in Experiment 3, focusing on star-galaxy classification performance. SENet-MargNet achieves high accuracy rates of 88.48$\%$ for classifying galaxies and 98.43$\%$ for stars. Importantly, misclassifications between stars and galaxies are minimal across all models, with SENet-MargNet, MargNet, ViT-MargNet, and MM ViT achieving rates as low as approximately 6.54$\%$, 6.85$\%$, 7.39$\%$, and 7.41$\%$, respectively.}
    \label{fig:ex3_sg_cm}
\end{figure*}

\begin{figure*}[htpb]
    \centering
    \begin{subfigure}[b]{0.48\linewidth}
        \centering
        \includegraphics[width=\linewidth]{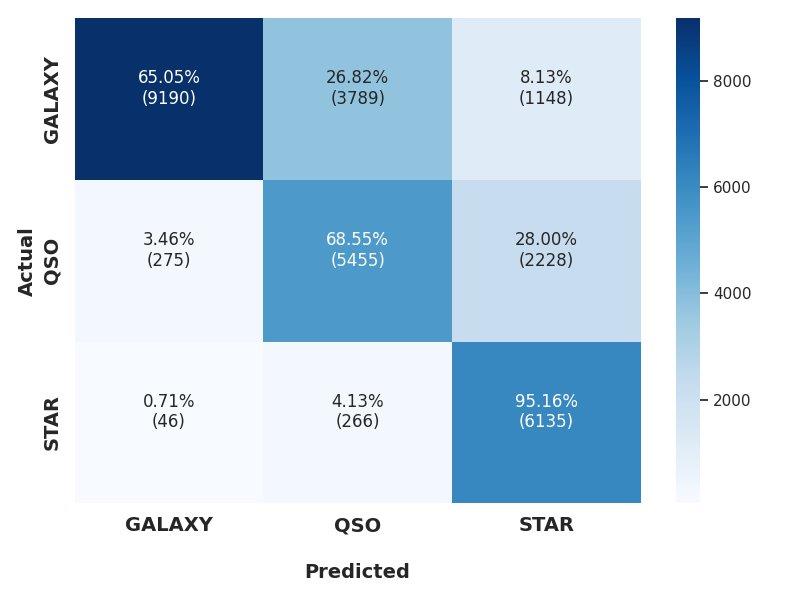}
        \caption{MargNet.}
        \label{fig:ex3_sgq_mgn}
    \end{subfigure}
    \hfill
    \begin{subfigure}[b]{0.48\linewidth}
        \centering
        \includegraphics[width=\linewidth]{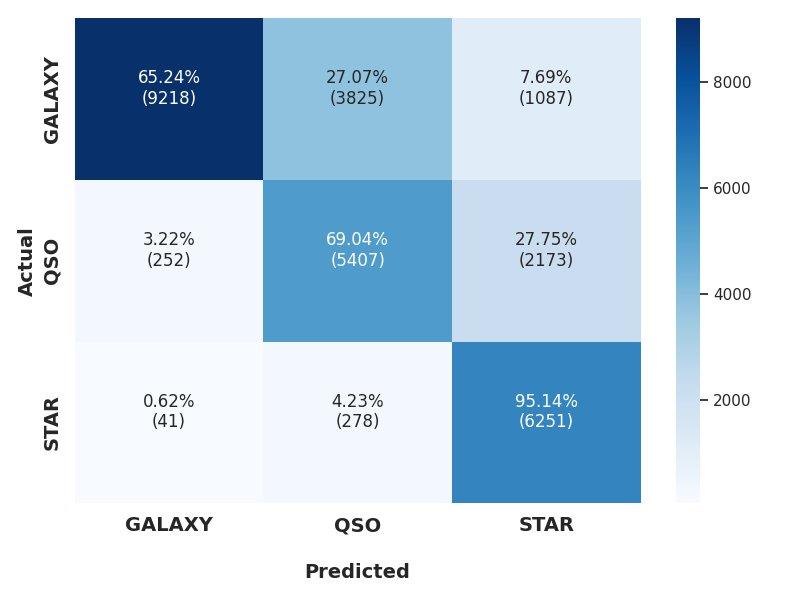}
        \caption{SENet-MargNet.}
        \label{fig:ex3_sgq_attmgn}
    \end{subfigure}
    
    \vskip\baselineskip
    
    \begin{subfigure}[b]{0.48\linewidth}
        \centering
        \includegraphics[width=\linewidth]{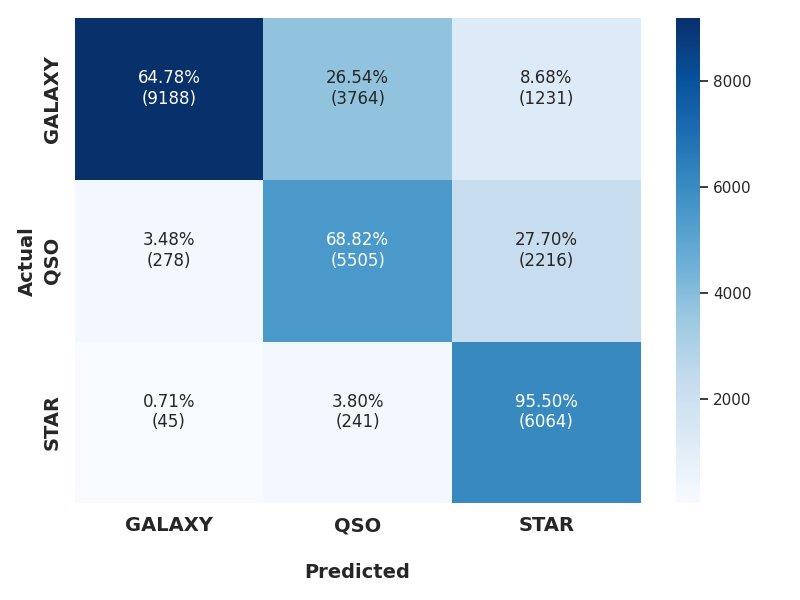}
        \caption{ViT-MargNet.}
        \label{fig:ex3_sgq_vitmgn}
    \end{subfigure}    
    \hfill
    \begin{subfigure}[b]{0.48\linewidth}
        \centering
        \includegraphics[width=\linewidth]{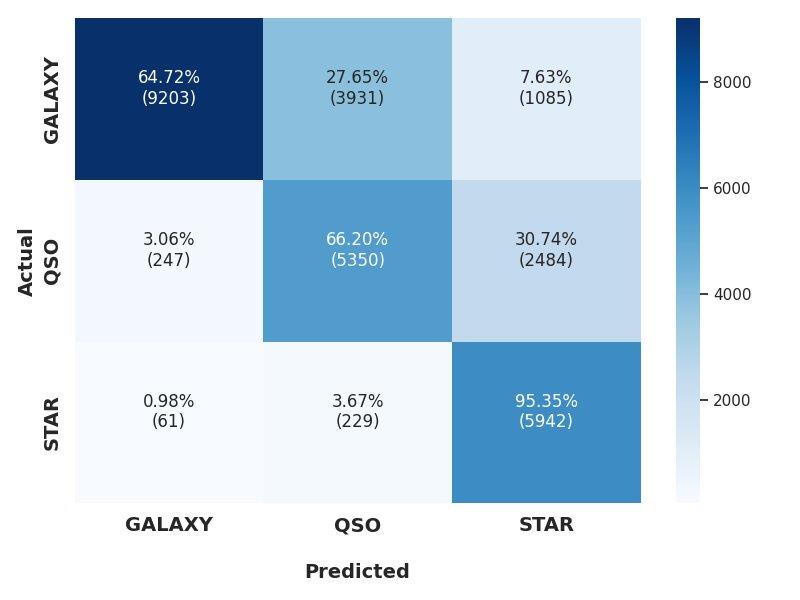}
        \caption{MM ViT.}
        \label{fig:ex3_sgq_mmvitmgn}
    \end{subfigure}
    
    \caption{Confusion matrices comparing MargNet (a), SENet-augmented MargNet (b), ViT-MargNet (c) and MM ViT (d) for Experiment 3 reveal their performance in star-galaxy-quasar classification. SENet-MargNet achieves 65.24$\%$ accuracy for galaxies, 69.04$\%$ for quasars, and 95.14$\%$ for stars. Notably, misclassifications among stars, quasars, and galaxies are minimal, with SENet-MargNet, MargNet, ViT-MargNet and MM ViT achieving error rates as low as approximately 11.76$\%$, 11.87$\%$, 11.81$\%$, and 12.28$\%$ respectively.}
    \label{fig:ex3_sgq_cm}
\end{figure*}

Experiment 3 selects the training and validation sets from the Compact Source dataset, while the test set is drawn from the Faint and Compact Source dataset. As in Experiments 1 and 2, a range of models are trained and evaluated using the Experiment 3 test dataset, and their performance metrics are summarized in Table \ref{tab:results_images_pf}. Given that the training data remains consistent between Experiments 1 and 3, we directly employ the models trained on Experiment 1's training data to assess their performance on the Experiment 3 test data. However, across all metrics, a decline in performance is evident compared to Experiment 1, attributed to the lower SNR as objects become fainter. For star-galaxy separation, SENet-MargNet achieves an accuracy of 92.9 $\pm$ 0.1$\%$, while MargNet, ViT-MargNet, and MM ViT achieve 92.0$\pm$0.1$\%$, 91.7$\pm$0.1$\%$, and 91.8$\pm$0.1$\%$ accuracy, respectively. The confusion matrices illustrated in Figure \ref{fig:ex3_sg_cm} highlight SENet-MargNet's excellence with an accuracy of 88.48$\%$ for galaxies and 98.43$\%$ for stars, followed closely by MargNet with accuracies of 87.67$\%$ for galaxies and 98.62$\%$ for stars. ViT-MargNet and MM ViT achieve accuracies of 86.75$\%$ and 86.81$\%$ for galaxies and 98.46$\%$ and 98.37$\%$ for stars, respectively. Notably, minimal misclassifications between stars and galaxies are observed, with rates of approximately 6.54$\%$, 6.85$\%$, 7.39$\%$, and 7.41$\%$ for SENet-MargNet, MargNet, ViT-MargNet, and MM ViT, respectively.

Similarly, for the star-galaxy-quasar classification, SENet-MargNet achieves an average accuracy of 73.2 $\pm$ 0.2$\%$, on par with MargNet, ViT-MargNet, and MM ViT, which achieve accuracies of 73.1$\pm$0.2$\%$, 72.7$\pm$0.2$\%$, and 71.8$\pm$0.2$\%$, respectively. Figure \ref{fig:ex3_sgq_cm} illustrates the individual class accuracies. Quasars exhibit the lowest accuracy, ranging from 66.2$\%$ with MM ViT to 69.04$\%$ with SENet-MargNet. For stars, SENet-MargNet achieves the lowest accuracy at 95.14$\%$, while ViT-MargNet achieves the highest at 95.5$\%$. Similarly, in galaxies, the lowest accuracy of 64.72$\%$ is attained using MM ViT, while SENet-MargNet achieves the highest at 65.24$\%$. Notably, misclassifications among stars, quasars, and galaxies remain minimal, with SENet-MargNet, MargNet, and MM ViT demonstrating error rates as low as approximately 11.76$\%$, 11.87$\%$, 11.81$\%$, and 12.28$\%$, respectively. Across all experiments and classification settings, the proposed SENet-MargNet consistently outperforms all other models. While the performance enhancement achieved with SENet augmentation on MargNet incurs an additional computational cost of approximately 0.1$\%$ to 3$\%$ in trainable parameters, integrating ViT into MargNet and MM ViT requires only 3.4$\%$ and 5.8$\%$ of the parameters of the baseline MargNet model, respectively. Notably, MM ViT performs comparably to ViT-MargNet.

\section{Conclusions and Future Work}
\label{sec:conclusion}
Our study presents the results aimed at enhancing the performance of the previously introduced MargNet model, utilizing attention mechanism and ViT-based architectures for classifying stars, galaxies, and quasars in SDSS photometric images. Within the SDSS dataset, our proposed MargNet variants, particularly SENet-MargNet, exhibit superior classification accuracy compared to prior neural network-based approaches, particularly for compact and faint galaxies. Additionally, our ViT-MargNet and MM ViT models offer lightweight alternatives. A significant advantage of CNN and ViT-based models is their ability to learn useful features directly from images automatically. This alleviates the need for separate feature engineering, as typically required in traditional machine learning algorithms. Notably, a portion of MargNet's architecture incorporates inception net modules, which have demonstrated promising results in various image classification tasks and have gained widespread adoption within the computer vision community. While conventional networks tend to exhibit reduced accuracy as sources become fainter and more compact, SENet-MargNet showcases the promising performance, achieving an overall accuracy of 93.5$\%$, with MargNet achieving 93.5$\%$, and ViT-MargNet and MM ViT models achieving accuracies of 93.2$\%$ each. When considering faint and compact sources, individual CNN and ANN components achieve approximately 91.6$\%$ and 93.0$\%$ accuracy, respectively. However, notable improvement in accuracy is observed when these components are combined within MargNet and MM ViT. The applicability of MargNet-based models can be extended to future surveys, given their ability to effectively capture numerous faint and compact sources. Furthermore, MargNet holds promise for surveys like GAIA and ZTF, which rely on crossmatching with SDSS and employ transfer learning techniques. Even in scenarios where photometric features differ from SDSS, adjustments to the ANN network within MargNet can accommodate alternative features. Alternatively, the image-based CNN component of MargNet can be utilized independently if photometric features are unavailable. In conclusion, the methodology underlying MargNet proves valuable due to its high accuracy in object classification, making it well-suited for deployment in forthcoming astronomical surveys with deeper observational reach.

\section{Acknowledgements}
Srinadh Reddy was supported by Tata Consultancy Services (TCS) and the Department of Science and Technology - Interdisciplinary Cyber-Physical Systems (DST-ICPS) (under the grant T-641). He is grateful to Ajit Kembhavi and Yogesh Wadadekar for useful feedback on his thesis defense presentation.
We thank the Sloan Digital Sky Survey for making their data free and open source. The Alfred P. Sloan Foundation, the U.S. Department of Energy Office of Science, and the Participating Institutions have funded SDSS IV. SDSS-IV acknowledges support and resources from the Center for High-Performance Computing at the University of Utah. The SDSS website is www.sdss.org. SDSS-IV is managed by the Astrophysical Research Consortium for the Participating Institutions of the SDSS Collaboration including the Brazilian Participation Group, the Carnegie Institution for Science, Carnegie Mellon University, Center for Astrophysics | Harvard $\&$ Smithsonian, the Chilean Participation Group, the French Participa- tion Group, Instituto de Astrofísica de Canarias, The Johns Hopkins University, Kavli Institute for the Physics and Mathematics of the Uni- verse (IPMU) / University of Tokyo, the Korean Participation Group, Lawrence Berkeley National Laboratory, Leibniz Institut für Astro- physik Potsdam (AIP), Max-Planck-Institut für Astronomie (MPIA Heidelberg), Max-Planck-Institut für Astrophysik (MPA Garching), Max-Planck-Institut für Extraterrestrische Physik (MPE), National Astronomical Observatories of China, New Mexico State Univer- sity, New York University, University of Notre Dame, Observatário Nacional / MCTI, The Ohio State University, Pennsylvania State University, Shanghai Astronomical Observatory, United Kingdom Participation Group, Universidad Nacional Autónoma de México, University of Arizona, University of Colorado Boulder, University of Oxford, University of Portsmouth, University of Utah, Univer- sity of Virginia, University of Washington, University of Wisconsin, Vanderbilt University, and Yale University.

\section{Software}
Astropy ~\citep{robitaille2013astropy}, Numpy ~\citep{harris2020array}, Scipy ~\citep{virtanen2020scipy}, Matplotlib ~\citep{hunter2007matplotlib}, Seaborn ~\citep{waskom2021seaborn}, Pandas ~\citep{reback2022pandas}, Jupyter ~\citep{kluyver2016jupyter}, scikit-learn ~\citep{pedregosa2011scikit}, Python3 ~\citep{van2009python}, and Pytorch ~\citep{paszke2019pytorch}.

\begin{appendices}

\section{Model Training Details}
We have implemented our deep-learning models in PyTorch 1.9.0 using the Adam optimizer \citep{kingma2014adam}. The CNN, SENet-CNN, and ViT (for images) and ANN (for photometric features) models underwent training with an early stopping criterion, ceasing training if there was no improvement over 30 and 100 consecutive epochs, respectively. Finally, the ensemble models (MargNet, SENet-MargNet and ViT-MargNet) were trained for 100 epochs for the last layers freezing the corresponding photometric feature-based and image-based models, with the best-performing models being saved. For the MM ViT, an early stopping criterion was applied, ceasing training when there was no further improvement over 100 consecutive epochs. Each model (ANN, CNN and ViT-based) was trained five times during training, and the average results are presented.

\section{Data Availability}
The PyTorch implementation of the proposed framework has been made open-source and is publicly available on \href{https://github.com/srinadh99/Astronomical-Source-Separation-using-Machine-Learning}{GitHub}. Photometric data was acquired from SDSS \href{https://skyserver.sdss.org/casjobs/}{Casjobs} DR16 using a specific query, while image data was obtained from \href{https://data.sdss.org/sas/dr16/eboss/photoObj/}{SDSS} through a Python script. The fully preprocessed dataset is accessible on \href{https://zenodo.org/records/6659435}{Zenodo}.




\end{appendices}

\bibliographystyle{sn-basic}
\bibliography{sn-bibliography}

\end{document}